\documentclass[twocolumn,superscriptaddress,longbibliography,prc]{revtex4-2}
\usepackage{times}
\usepackage[T1]{fontenc}
\usepackage{CJKutf8}

\usepackage{pdfpages}
\usepackage{pgffor}

\makeatletter
\AtBeginDocument{\let\LS@rot\@undefined}
\makeatother

\usepackage{physics}
\usepackage{amsmath,amssymb,mathrsfs}
\usepackage{graphicx}
\usepackage{dcolumn}
\usepackage{bm}
\usepackage{morefloats}
\usepackage{multirow}
\usepackage{amssymb}
\usepackage{amsmath}
\usepackage{xcolor}
\usepackage{longtable}
\usepackage{fix-cm}
\usepackage{verbatim}
\usepackage{float}
\usepackage{mathptmx} 
\usepackage[T1]{fontenc}
\usepackage[colorlinks,allcolors=blue]{hyperref}

\usepackage[normalem]{ulem}

\makeatletter
\def\NAT@def@citea{\def\@citea{\NAT@separator}}
\makeatother

\def\0\\{\nonumber\\}

\newcommand{\beq}{\begin{align}}
\newcommand{\eeq}{\end{align}}
\newcommand{\beqn}{\begin{align}}
\newcommand{\eeqn}{\end{align}}

\makeatletter
\newcommand\footnoteref[1]{\protected@xdef\@thefnmark{\ref{#1}}\@footnotemark}
\makeatother

\begin{document}

\title{
Microscopic description of spin distributions in multinucleon tranfer reactions
}

\author{Kotaro Koga}
\email{koga.k.c19d@m.isct.ac.jp}
\affiliation{Department of Physics, School of Science, Institute of Science Tokyo, Tokyo 152-8550, Japan}

\author{Kazuyuki Sekizawa}
\email{sekizawa@phys.sci.isct.ac.jp}
\affiliation{Department of Physics, School of Science, Institute of Science Tokyo, Tokyo 152-8550, Japan}
\affiliation{Nuclear Physics Division, Center for Computational Sciences, University of Tsukuba, Ibaraki 305-8577, Japan}
\affiliation{RIKEN Nishina Center, Saitama 351-0198, Japan}

\date{\today}

\begin{abstract}

\edef\oldrightskip{\the\rightskip}
\begin{description}
\rightskip\oldrightskip\relax
\setlength{\parskip}{0pt} 
\item[Background]
The generation of spin distributions in fission fragments has attracted substantial interest in recent
years and its mechanism is becoming clear alongside the refined description with microscopic theories.
In contrast, the microscopic origin of spin distributions in reaction products remain largely unexplored,
although it is not only an important input for modeling their statistical deexcitation but also carries
essential information to understand the underlying reaction mechanism.

\item[Purpose]
The primary purpose of this work is to demonstrate to what extent the time-dependent Hartree-Fock (TDHF)
theory can describe spin distributions of reaction products in respective $(N,Z)$ transfer channel
and how much it deepens our understanding of the reaction mechanism.

\item[Methods]
Three-dimensional TDHF calculations are performed for multinucleon transfer processes in the
$^{40}\text{Ca} + {}^{124}\text{Sn}$ reaction at $E_{\text{c.m.}} \simeq 128.54\,\text{MeV}$ using
the Skyrme SLy5 energy density functional. To obtain spin distributions in each transfer
channel, the particle-number projection (PNP) and the angular-momentum projection (AMP) are simultaneously
achieved for a reaction product.

\item[Results]
By performing PNP+AMP for TDHF wave functions after collisions, spin-dependent transfer cross sections
are calculated for projectile- ($^{40}$Ca-) like fragments (PLFs). From the results, we find that
abundant spin distributions up to $J=20\hbar$ or more can be obtained from the TDHF wave functions.
For one-nucleon addition and removal channels, the calculated spin values agree well with those of
single-particle states in the $pf$- and $sd$-shell, respectively. Moreover, for a-few-nucleon transfer
channels, we show that the obtained spin distributions can be well explained by combinations of total
angular momenta of those single-particle states. By carefully investigating the fragment spin, we can infer
to which state and with what probability the transfer occured, the information which is never accessible
without the PNP+AMP analysis. Furthermore, as the number of transferred nucleons increases, we find that
the calculated spin distributions exhibit a smooth singly-peaked shape with a long tail as a function
of $J$, which resembles that of empirical spin distributions for thermally equilibrated fragments.
An exploratory calculation for induced fission of $^{240}$Pu is also carried out,
obtaining similar results as in multinucleon transfer reactions.

\item[Conclusions]
It is shown that TDHF is capable of describing wide, abundant spin distributions in
multinucleon transfer reactions near the Coulomb barrier, which substantially deepens
our understanding of the unerlying reaction mechanism. The successful description is achieved,
because the spin distributions arise mainly from complex combinations of single-particle
angular momenta and there are a sufficient number of states even in the single mean-field
potential in the TDHF theory.

\end{description}
\end{abstract}
\maketitle

\section{Introduction}\label{Sec:Intro}

Multinucleon-transfer (MNT) reactions at energies near the Coulomb barrier have been
extensively investigated at experimental facilities over the world~\cite{Mijatovic_2022,
Watanabe_2015,Kozulin_2012,Leguillon_2016,Vermeulen_2020,Caamano_2013,Szilner_2005,
Corradi_1996,Barrett_2015,Desai_2019,Kozulin_2024,Szilner_2007,Watanabe_2021,Ahmed_2021}.
These studies have pursued a variety of objectives, including the production of neutron-rich
nuclei that are difficult to access through other reaction mechanisms~\cite{Watanabe_2015,Kozulin_2012},
investigations of transfer-induced fission~\cite{Leguillon_2016,Vermeulen_2020,Caamano_2013},
elucidation of reaction mechanisms~\cite{Corradi_1996,Barrett_2015,Desai_2019,Kozulin_2024},
and studies of nuclear structure~\cite{Szilner_2007,Watanabe_2021,Ahmed_2021}.

The MNT reaction occurs, in general, in the grazing collision at nonzero initial orbital
angular momenta, part of which can be converted into the intrinsic angular momenta of the
outgoing reaction products. The angular-momentum distributions of the primary fragments
not only contain precious information to disentangle the complex reaction mechanism of
MNT reactions, but also important as it affects observables measured after their deexcitation.
For instance, the isomeric yield ratio can depend sensitively on the initial angular-momentum
distributions of the reaction products~\cite{Kumar_2024}. Despite the apparent significance of
the angular-momentum distributions in low-energy heavy-ion reactions, their miscroscopic
description has not been well explored todate. In this work, we attack this problem with
the time-dependent Hartree--Fock (TDHF) theory \cite{Dirac_1930,Negele_1982,Simenel_2018,Sekizawa_2019_1}
and show that abundant angular-momentum distributions can be obtained even with the
single-Slater description of the TDHF theory.

Traditionally, fragment spin has been attributed to the ``sliding'' and ``rolling'' modes
within macroscopic or semiclassical frameworks~\cite{Tsang_1974,Bondorf_1974}. Treating
colliding nuclei as rigid bodies, the sliding mode denotes motion in which their surfaces
slip relative to each other at the point of contact, whereas the rolling mode denotes
motion in which they remain in contact without slipping. However, from a microscopic
perspective, nuclei are not rigid bodies but are composed of neutrons and protons that
interact through the nuclear force, and even the definition of their surfaces is
ambiguous because the surfaces evolve dynamically in the course of collision.
Therefore, fully-microscopic, dynamic description is desired to elucidate the
generation mechanism of the angular-momentum distributions of reaction products
in low-energy heavy-ion reactions.

To study MNT reactions, a variety of phenomenological, semiclassical, and transport
approaches have been developed and successfully applied, including: the dinuclear-system
(DNS) model~\cite{Adamian_1997,Adamian_2010,Zhu_2017,Feng_2017,Guo_2019,Bao_2022}, GRAZING~\cite{Winther_1994,Winther_1995,Corradi_2002}, complex
Wentzel--Kramers--Brillouin (CWKB)~\cite{Vigezzi_1989,Corradi_2002}, a dynamical model with
Langevin-type equations of motion~\cite{Zagrebaev_2005,Zagrebaev_2008,Saiko_2019,Saiko_2022}, and improved quantum molecular
dynamics (ImQMD) model~\cite{Wang_2002,Li_2016,Yao_2017}.
These approaches generally involve phenomenological ingredients or model parameters
constrained by available experimental data. On the other hand, microscopic TDHF
calculations require no reaction-specific adjustment of model parameters, although
the EDF itself is to some extent phenomenological and leaves a room for further
modifications and refinements. Once the EDF, initial nuclear states and collision conditions
are specified, the subsequent dynamics are determined self-consistently. The TDHF theory
with a Skyrme-type effective interaction may be regarded as an incarnation of the
time-dependent density functional theory (TDDFT)~\cite{Runge_1984,Nakatsukasa_2016},
including extensions and related formulations, such as time-dependent Hartree--Fock--Bogoliubov
(TDHFB) theory or time-dependent superfluid local density approximation (TDSLDA)~\cite{Bulgac_2016},
which incorporates pairing correlations, and time-dependent covariant density functional theory
(TDCDFT)~\cite{Ren_2020}, which provides a relativistic formulation. Owing to their
self-consistent character, TDHF and related TDDFT approaches have been applied not
only to MNT reactions~\cite{Sekizawa_2013,Simenel_2010,Scamps_2024,Zhang_2025,Wu_2019}
but also to a wide variety of nuclear dynamics, including fusion~\cite{Umar_2006_2,
Umar_2006_1,Umar_2010,Guo_2018,Sekizawa_2019_2}, quasifission~\cite{Golabek_2009,
Washiyama_2015,Umar_2015,Sekizawa_2016,McGlynn_2023}, induced fission~\cite{Scamps_2018,
Simenel_2014,Goddard_2015,Goddard_2016,Marevic_2026}, and linear responses
\cite{Nakatsukasa_2005,Maruhn_2005,Ebata_2010,Scamps_2013,Ebata_2014}.

Naively, the mechanism of nucleon transfer was investigated by analyzing
distributions of single-particle wave functions after collision, which
were initially identified as specific orbitals in projectile or target
nuclei~\cite{Umar_2008}. Such an analysis of single-particle wave functions
may provide a hint on the reaction mechanism, however, it is not well defined,
because an arbitrary unitary transformation among the occupied
single-particle orbitals, $\phi_i'(\mathbf{r}\sigma)=\sum_jU_{ij}
\phi_i(\mathbf{r}\sigma)$, leaves the Slater determinant unchanged
(except a redundant phase factor), \textit{i.e.} $\Phi'= e^{i\delta}\Phi$,
where $e^{i\delta}=\det U$ with an arbitrary phase $\delta$. Contrary,
the projection technique offers a way to analyze the TDHF wave function
after collision in a gauge invariant manner, as a quantity like
$\mathcal{O}_P=\bigl<\Phi\big|\hat{\mathcal{O}}\hat{P}\big|\Phi\bigr>$,
with an artbitrary projection operator $\hat{P}$, is apparently not
affected by the unitary transformation, where $\hat{\mathcal{O}}$ is
an arbitrary operator.

Indeed, projection methods
have played an important role in TDHF studies of MNT reactions. A TDHF wave
function after collision is, in general, a superposition of eigenstates
of observables such as particle numbers and angular momenta. Accordingly,
the particle-number projection (PNP) method~\cite{Simenel_2010,Sekizawa_2013}
enables the extraction of probability distributions for specific proton numbers
$Z$ and neutron numbers $N$. Numerous successful applications of PNP have
been reported. In Ref.~\cite{Sekizawa_2014}, the PNP method has been extended
to calculate the expectation value of operators, such as energy and angular
momentum, in the particle-number projected state. Although a careful analysis
of those expectation values provides a clue for identifying the transfer
mechanism, the analysis of Ref.~\cite{Sekizawa_2014} was restricted to
the expectation value with PNP, \textit{i.e.}, without the angular-momentum
projection (AMP). One may investigate the angular-momentum generation
mechanism in heavy-ion reactions by analyzing the expectation values as
was successfully done in, \textit{e.g.}, Ref.~\cite{Scamps_2024}. Such
an analysis allows us to understand the main, averaged dynamics during
the collision. However, to access full information about angular-momentum
distributions in reaction products, we do need to use AMP for TDHF wave
functions after collision.

The AMP method~\cite{Ring_1980,Bulgac_2021,Marevic_2026} provides a powerful
means of determining amplitudes of each angular-momentum state, which
has originally been used for nuclear structure studies, \textit{e.g.},
for symmetry restoration of deformed mean-field solutions or of
generator-coordinate-type superposition of various wave functions
\cite{Ring_1980,Bender_2003}. The application of AMP to fragmented
wave functions was pioneered by Ref.~\cite{Scamps_2023}, in the context
of induced fission, where angular-momentum distributions
of fission fragments were calculated by applying AMP within TDSLDA,
and similar AMP studiy within TDCDFT was reported in Ref.~\cite{Li_2026}.
More recently, PNP, AMP, as well as the parity projection have been applied
simultaneously within TDHFB and angular-momentum distributions for natural
and unnatural parity states were calculated for induced-fission fragments
\cite{Scamps_2026}. In Ref.~\cite{Zhang_2025}, the AMP method was actually
applied for MNT reactions, but PNP was not performed. As a result,
the distributions were not resolved into individual $(N,Z)$ channels.

In this study, we simultaneously apply PNP and AMP to each reaction fragment
in the TDHF wave functions after collision. Whereas in the preceding study
of Ref.~\cite{Zhang_2025} the mechanisms of spin generation and quantum
entanglement were investigated, here we focus specifically on primary-fragment
angular-momentum distributions in each transfer channel. We investigate the
$^{40}\text{Ca} + {}^{124}\text{Sn}$ reaction which was studied both
experimentally and theoretically. We determine the spin-dependent transfer
cross sections by calculating the joint probability for $(N,Z,J)$ at each
impact parameter $b$. From these cross sections, we derive the primary-fragment
angular-momentum distributions for projectile-like ($^{40}$Ca-like) fragments
specified by $(N,Z)$. We then discuss the physical insight provided by these
distributions, which tells us how the MNT reaction mechanism is altered
depending on the impact parameter and the number of transferred nucleons.
In addition, to demonstrate the usefulness of our PNP+AMP combined
analysis, we perform expolatory analysis of angular-momentum distributions
in fission fragments in induced fission of $^{240}$Pu.

The article is organized as follows. In Sec.~\ref{Sec:Methods}, the TDHF
framework is succinctly recalled and the PNP and AMP methods used to obtain
fragment-resolved angular-momentum distributions and spin-dependent transfer
cross sections are introduced. After providing the computational setup in
Sec.~\ref{Sec:Details}, the calculated spin-dependent transfer cross
sections and angular-momentum distributions are presented and discussed
in Sec.~\ref{Sec:Results}. Finally, a summary and perspectives
are given in Sec.~\ref{Sec:Conclusion}.

\section{Methods}
\label{Sec:Methods}

\subsection{Time--Dependent Hartree-Fock (TDHF) Theory}

In the TDHF theory, the many-body wave function is represented by
\begin{equation}
    \ket{\Phi(t)} = \bigl|\Phi^{(\text{n})}(t)\bigr>\bigl|\Phi^{(\text{p})}(t)\bigr>,
\end{equation}
where $\bigl|\Phi^{(q)}\bigr>$ is a single Slater determinant for neutrons
($q=\text{n}$) or protons ($q=\text{p}$),
\begin{align}
 \ket{\Phi(t)}
 &=
 \frac{1}{\sqrt{N_q!}}\det[\phi_i^{(q)}(\mathbf{r}_j\sigma_j)].
 \label{eq:slater}
\end{align}
Here, $\sigma=\pm 1/2$ denotes the spin quantum number. The occupied single-particle
wave functions are orthonormal to each other for all times. Their time evolution is
governed by the TDHF equation,
\begin{align}
 i\hbar\frac{\partial}{\partial t}
 \phi_i(\mathbf r,\sigma,q;t)
 &=
 \sum_{\sigma'}\hat h_{\sigma\sigma'}^{(q)}(\mathbf{r},t)\phi_i^{(q)}(\mathbf r,\sigma',t),
 \label{eq:tdhf}
\end{align}
where $\hat h_{\sigma\sigma'}^{q}(t)$ is the single-particle Hamiltonian that contains
mean-field potentials which are dependent on one-body densities. The ground-state
single-particle wave functions of projectile and target nuclei are obtained separately
by solving the static Hartree--Fock equations,
\begin{align}
\sum_{\sigma'} \hat h_{\sigma\sigma'}^{(q)}(t_i)\phi_i^{(q)}(\mathbf r,\sigma',t_i)
 &=
 \varepsilon_i^{(q)}\phi_i(\mathbf r,\sigma,t_i).
 \label{eq:static_hf}
\end{align}
The projectile and target nuclei are placed in a computational box with a certain
relative distance, where the relative momentum is evaluated by assuming the Rutherford
trajectory for a given set of incident energy $E_{\mathrm{c.m.}}$ and impact parameter $b$.
Time evolution is calculated until the binary reaction products are well separated
at the final time of the simulation, $t_\text{f}$.

\subsection{Projection technique}

We analyze a TDHF wave function after collision at $t_\text{f}$ using PNP and AMP.
In the following, we express the TDHF wave function at the final time $t_\text{f}$
as $\ket{\Psi}=\ket{\Phi(t_\text{f})}$. Similarly, single-particle wave functions
at $t_\text{f}$ are represented as $\psi_i^{(q)}(\mathbf{r}\sigma)=
\phi_i^{(q)}(\mathbf{r}\sigma,t_\text{f})$.

\subsubsection{Particle-number projection (PNP)}

At the initial time of the simulation, the total TDHF wave function is solely
a direct product of HF wave functions which are eigenstate of the particle-number
operator with eigenvalue $N_\text{P}$ and $Z_\text{P}$ for the projectile and
$N_\text{T}$ and $Z_\text{T}$ for the target. On the course of the reaction,
nucleons can be exchanged between colliding nuclei and, as a result, single-particle
wave functions can, in general, be distributed in both a projectilelike fragment
(PLF) and a targetlike fragment (TLF). Because of this fact, while the TDHF
wave function remains an eigenstate of the particle-number operator in the
whole space with eigenvalues $N_\text{P}+N_\text{T}$ and $Z_\text{P}+Z_\text{T}$,
it is not an eigenstate of the particle-number operator in a subspace $V$
which involves one of the reaction products. Therefore, the TDHF wave function
after collision can be regarded as a superposition of states with different
particle number distributions (or transfer channels).

To extract the probability for producing a rection product specified by
a set of neutron and proton numbers, $(N,Z)$, we use the PNP method.
The PNP operator that projects the TDHF wave function onto a state
which contains $n$ particles in a spatial region $V$ is defined as
\begin{align}
 \hat P_n^{(q)}
 &=
 \frac{1}{2\pi}\int_0^{2\pi}\,
 e^{i\theta(\hat N_V^{(q)}-n)}\dd\theta,
 \label{eq:pnp}
\end{align}
where $n$ is an arbitrary integer number and $\theta$ is called the
gauge angle. Here, $N_V^{(q)}$ denotes the particle-number operator
for the spatial region $V$, which is defined as
\begin{align}
 \hat N_V^{(q)}
 &=
 \sum_{i\in q}\int 
 \Theta_V(\mathbf r)\delta(\mathbf{r}-\hat{\mathbf{r}}_i)\dd\mathbf{r},
 \label{eq:local_number}
\end{align}
where $\Theta_V(\mathbf r)$ is a Heviside step function which is unity inside $V$
and zero elsewhere.

In ordinary TDHF calculations that do not allow proton--neutron mixing,
the TDHF wave function after collision is a direct prouct of wave functions
for neutrons and protons, $\ket{\Psi}=\ket{\Psi_n}\ket{\Psi_p}$. In such a
case as studied here, the probability of producing a reaction product
composed of $N$ neutrons and $Z$ protons in the volume $V$ is given by
a product of the probabilities for neutrons and protons,
\begin{align}
 P(N,Z)
 &=
 \mel{\Psi}{\hat P_N^{(n)}\hat P_Z^{(p)}}{\Psi}
 \notag\\
 &=
 \mel{\Psi_n}{\hat P_N^{(n)}}{\Psi_n}
 \mel{\Psi_p}{\hat P_Z^{(p)}}{\Psi_p}
 \notag\\
 &=
 P_N^{(n)}P_Z^{(p)}.
 \label{eq:pnp_joint}
\end{align}
where
\begin{align}
 P_n^{(q)}
 &=
 \frac{1}{2\pi}\int_0^{2\pi}
 e^{-in\theta}
 \det\!\left[
 \delta_{ij}-S_{V,ij}^{(q)}+e^{i\theta}S_{V,ij}^{(q)}
 \right]\dd\theta.
 \label{eq:pnp_single}
\end{align}
The determinant in Eq.~\eqref{eq:pnp_single} is taken over the occupied
single-particle states with a specific $q$. $S_{ij}^{(q)}$ denotes the
$ij$-component of the overlap matrix evaluated in the volume $V$,
\begin{align}
 S_{V,ij}^{(q)}
 &=
 \sum_{\sigma}\int_V
 \psi_i^{(q)*}(\mathbf r,\sigma)
 \psi_j^{(q)}(\mathbf r,\sigma) \dd\mathbf{r}.
 \label{eq:pnp_overlap}
\end{align}

The PNP method has been widely used to calculate transfer probabilities
and transfer cross sections which can be directly compared with experimental
data [...]. In the preceeding studies, it was shown that TDHF provides
quantitative description of transfer cross sections for main reaction
channels. In those studies, transfer cross sections were inclusive,
in the sense that the quantum states (spin-parity and excitatoin energy)
of reaction products are not resolved, but averaged out.

\subsubsection{Angular-momentum projection (AMP)}

On the course of the reaction, part of the collision energy of relative motion
dissipates into internal degrees of freedom, not only single-particle excitations,
but also shape deformation and collective surface vibrations and rotation. Thus,
the single-particle wave functions in the TDHF wave function after collision are,
in general, complex mixture of excited states and the part of the TDHF wave function
for a reaction product in the volume $V$ can be regarded as a complex superposition
of various excitated nuclear states.

As a first step to calculate exclusive cross sections, here we consider the AMP method.
The AMP operator that projects the TDHF wave function onto a state which has total angular
momentum $J$, summed over its magnetic substates, is defined as
\begin{align}
 \hat P_J
 &=
 \frac{2J+1}{16\pi^2}
 \int_0^{2\pi}\!\dd\alpha
 \int_0^\pi\!\dd\beta\,\sin\beta
 \int_0^{4\pi}\!\dd\gamma\;
 \chi_J^*(\Omega)\hat R_V(\Omega),
 \label{eq:amp}
\end{align}
where $\Omega$ represents the Euler angle, $\Omega=(\alpha,\beta,\gamma)$,
$\chi_J(\Omega)=\sum_{M=-J}^{J}D^J_{MM}(\Omega)$ is the character of the $J$
representation in terms of the Wigner $D$ functions, and $\hat{R}_V(\Omega)$
is a rotation operator for the volume $V$. The range $0\leq\gamma<4\pi$
covers the rotation group in its $\mathrm{SU}(2)$ representation and therefore
accommodates both integer and half-integer values of $J$. The rotation operator
is defined as
\begin{align}
 \hat R_V(\Omega)
 &=
 \exp[-i\alpha\hat J_{V,z}/\hbar]
 \exp[-i\beta\hat J_{V,y}/\hbar]
 \exp[-i\gamma\hat J_{V,z}/\hbar],
 \label{eq:local_rotation}
\end{align}
where $\hat{\mathbf J}_V=(\hat{J}_{V,x},\hat{J}_{V,y},\hat{J}_{V,z})$ is the
total angular-momentum operator in the volume $V$, including both the orbital
and intrinsic-spin contributions, defined as
\begin{align}
 \hat{\mathbf J}_V
 &=
 \Theta_V(\mathbf r)
 \sum_i
 \left[
 (\hat{\mathbf{r}}_i-\mathbf R_V)\times\hat{\mathbf p}_i
 +\hat{\mathbf s}_i
 \right]
 \label{eq:local_angular_momentum}
\end{align}
where $\hat{\mathbf p}=-i\hbar\boldsymbol{\nabla}$ and $\hat{\mathbf s}=(\hbar/2)\boldsymbol{\sigma}$
are the single-particle momentum and spin operators, respectively, where $\boldsymbol{\sigma}$
is the Pauli matrix in the spin space. The vector $\mathbf R_V=\int_V n(\mathbf{r})\mathbf{r}\,\dd\mathbf{r}
/\int_V n(\mathbf{r})\dd\mathbf{r}$ denotes the center-of-mass position of the reaction product
in the volume $V$. Hence, the orbital angular momentum and the spatial rotation are defined
with respect to the center-of-mass position of the selected PLF or TLF.

On a single-particle wave function, the rotation acts on both the spatial and spin coordinates according to
\begin{align}
 \hat R(\Omega)\psi_j^{(q)}(\mathbf r,\sigma)
 &=
 \sum_{\sigma'}
 D^{1/2}_{\sigma\sigma'}(\Omega)
 \psi_j^{(q)}(R^{-1}(\Omega)\mathbf r,\sigma'),
 \label{eq:single_particle_rotation}
\end{align}
where $\mathbf r$ is defined in the coordinate fixed to the center-of-mass position of
the reaction product in the volume $V$. The probability that the fragment has total angular
momentum $J$ is
\begin{align}
 P(J)
 &=
 \mel{\Psi}{\hat P_J}{\Psi},
 \label{eq:amp_probability}
\end{align}
which can be evaluated as follows:
\begin{align}
 P(J)
 &=
 \frac{2J+1}{16\pi^2}
 \int \chi_J^*(\Omega)
 \prod_{q=n,p}
 \det\!\left[
 \delta_{ij}-S_{\bar{V},ij}^{(q)}
 +B_{V,ij}^{(q)}(\Omega)
 \right]\dd\Omega,
 \label{eq:amp_determinant}
\end{align}
where the $ij$-component of the rotated overlap matrix evaluated in the volume V is defined as
\begin{align}
 B_{V,ij}^{(q)}(\Omega)
 &=
 \sum_\sigma\int_V d^3r\,
 \psi_i^{(q)*}(\mathbf r,\sigma)
 \hat R(\Omega)\psi_j^{(q)}(\mathbf r,\sigma).
 \label{eq:rotated_overlap}
\end{align}
In Eq.~\eqref{eq:amp_determinant}, the term $\delta_{ij}-S_{\bar{V},ij}^{(q)}$ represents
the unrotated contribution evaluated in the complement of $V$, whereas $B_{V,ij}^{(q)}(\Omega)$
represents the locally rotated contribution evaluated in $V$. The translational motion
of the selected fragment is removed before the projected kernels are evaluated,
as in Ref.~\cite{Sekizawa_2014}.

\subsubsection{The PNP + AMP method}

To calculate spin distributions in each transfer channel, we need to perform
both PNP and AMP simultaneously. By combining the PNP and AMP operators, the
joint probability that the reaction product is composed of $N$ neutrons and
$Z$ protons and has total angular momentum $J$ can be expressed as
\begin{align}
 P(N,Z,J)
 &=
 \mel{\Psi}{
 \hat P_N^{(\text{n})}
 \hat P_Z^{(\text{p})}
 \hat P_J
 }{\Psi}.
 \label{eq:joint_definition}
\end{align}
The local rotation conserves the neutron and proton numbers within $V$ and acts
on the neutron and proton parts independently. Consequently, $\hat P_J$ commutes
with $\hat P_N^{(\text{n})}$ and $\hat P_Z^{(\text{p})}$ and the three quantum numbers
($N,Z,J$) can be projected simultaneously. The joint probability can be calculated as
\begin{align}
 P(N,Z,J)
 &=
 \frac{2J+1}{16\pi^2}
 \int 
 \chi_J^*(\Omega)
 K^{(\text{n})}(N,\Omega)K^{(\text{p})}(Z,\Omega)\dd\Omega,
 \label{eq:joint_probability}
\end{align}
where, for $q=\text{n},\text{p}$,
\begin{align}
 K^{(q)}(n,\Omega)
 &=
 \frac{1}{2\pi}\int_0^{2\pi}
 e^{-in\theta}
 \det[M_{ij}^{(q)}(\theta,\Omega)]\dd\theta,
 \label{eq:projected_rotation_kernel}\\[2mm]
 M_{ij}^{(q)}(\theta,\Omega)
 &=
 \delta_{ij}-S_{\bar{V},ij}^{(q)}
 +e^{i\theta}B_{V,ij}^{(q)}(\Omega).
 \label{eq:kernel}
\end{align}
Thus, $K^{(q)}(n,\Omega)$ is the particle-number-projected rotational kernel
for $n$ neutrons (for $q=\text{n}$) or $n$ protons (for $q=\text{p}$) at a
fixed Euler angle $\Omega$. In the PNP+AMP method, the marginal distributions
can also be calculated as
\begin{align}
 P(N,Z) = \sum_J P(N,Z,J),\\
 P(J) = \sum_{N,Z}P(N,Z,J).
 \label{eq:joint_normalization}
\end{align}

\subsection{Spin-resolved transfer cross sections}

TDHF calculations of the multinucleon transfer reaction are performed for
a range of impact parameters, covering (quasi)elastic to inelastic reactions
leading to production of binary reaction products. Let us denote the joint
probability obtained from the final state $\ket{\Psi(b)}$ as $P(b;N,Z,J)$.
The spin-resolved transfer cross section for producing a primary reaction
product specified by $(N,Z,J)$ can be calculated by integrating over the
impact parameter as
\begin{align}
 \sigma(N,Z,J)
 &=
 2\pi\int P(b;N,Z,J)\,b\,\dd b,
 \label{eq:cross_section}
\end{align}
where $2\pi b\,\dd b$ is the area element resulting from azimuthal symmetry
about the beam axis. Using the relation of the marginal distributions of
Eq.~\eqref{eq:joint_normalization}, the spin-integrated transfer cross sections
can be expressed as
\begin{equation}
    \sigma(N,Z) = \sum_J \sigma(N,Z,J).
\end{equation}
These spin-integrated transfer cross sections were evaluated in the preceeding
studies.

Since the magnitude of the spin-resolved transfer cross sections depend
largely on the number of transferred nucleons, it is useful to define
a normalized angular-momentum distribution for each transfer channel
to investigate the detailed reaction mechanism. The normalized
angular-momentum distribution for a given transfer channel $(N,Z)$
is defined as
\begin{align}
 P(J\mid N,Z)
&\equiv
 \frac{\sigma(N,Z,J)}
 {\sigma(N,Z)}.
 \label{eq:conditional_j}
\end{align}
We note that both $\sigma(N,Z,J)$ and $P(J\mid N,Z)$ analyzed in this work
refer to primary reaction products before statistical deexcitation processes
take place.

\section{Computational Details}\label{Sec:Details}

We have newly developed parallel computational codes of static HF
calculations (both 3D and with axial symmetry), of dynamic TDHF
calculations, and of the PNP+AMP analysis of TDHF wave functions
after collision, written in \texttt{Fortran90} language. In HF and
TDHF calculations, the OpenMP parallelization is adopted for
loops over the orbital index. In the AMP analysis, the integrations
over Euler angles are parallelized for multi-GPU architechture
with MPI and OpenACC. Typically, for a single PNP+AMP calculation
it takes roughly four hours with 16 GPUs at the Miyabi-G supercomputer
(NVIDIA H100) at the University of Tokyo.

The reaction calculations are performed for the \(^{40}\mathrm{Ca}+{}^{124}\mathrm{Sn}\)
reaction at \(E_{\mathrm{c.m.}}\simeq 128.54~\mathrm{MeV}\). The same
reaction was previously studied in TDHF with PNP in Ref.~\cite{Sekizawa_2013},
and the present TDHF setup follows that work. For the Skyrme EDF, we employ
the SLy5 parametrization \cite{Chabanat_1998} without the center-of-mass correction
in both HF and TDHF calculations. The \(J^2\) terms (\textit{i.e.}, the
$\sum_{\mu\nu}(J_{\mu\nu})^2$ term), and all time-odd terms (\textit{i.e.},
the time-odd spin-orbit terms, $\boldsymbol{s}^2$ terms, and $\boldsymbol{s\cdot T}$
terms) are retained, except the $\mathbf{s}\boldsymbol{\cdot}\Delta\mathbf{s}$ term
that causes spin instability. Pairing correlations are neglected in the present study.

The static HF ground states of the projectile and target nuclei are calculated
separately  with \(26\times26\times26\) mesh points with a spacing of \(0.8~\mathrm{fm}\).
The obtained wave functions are placed in a bigger box with \(60\times60\times26\)
mesh points with the same mesh spacing. In both HF and TDHF calculations,
spatial derivatives are evaluated using the \(11\)-point finite-difference formula.

For a given set of collision energy $E_\text{c.m.}$ and impact parameter \(b\),
the initial relative momenta are evaluated assuming a Rutherford trajectory,
and proper Galilean boost phases are applied to the projectile and target single-particle
wave functions to simulate the reaction in the center-of-mass frame. TDHF calculations
are performed at \(22\) impact parameters over the range \(3.95~\mathrm{fm}\leq
b\leq10~\mathrm{fm}\), using a generally nonuniform impact-parameter mesh. The
spacing was \(1~\mathrm{fm}\) from \(b=10\) to \(8~\mathrm{fm}\), \(0.5~\mathrm{fm}\)
from \(b=8\) to \(7~\mathrm{fm}\), \(0.25~\mathrm{fm}\) from \(b=7\) to \(4~\mathrm{fm}\),
and \(0.01~\mathrm{fm}\) from \(b=4\) to \(3.95~\mathrm{fm}\). Fusion occurred for
\(b_\text{f}\leq3.94~\mathrm{fm}\), which is consistent with the value ($b_\text{f}\le
3.95$) reported in Ref.~\cite{Sekizawa_2013}.

To calculate the time evolution in TDHF, the fifth-order Taylor expansion is
employed with a single predictor--corrector step with a time step of
\(\Delta t=0.2~\mathrm{fm}/c\) to approximately express the unitary
time-evolution operator. The time evolution is stopped when the binary
reaction products are well separated spatially, but not too close to the
edge of the computational box.

For the PNP and AMP calculations, the fragment region \(V\) was defined as a sphere
centered at the center-of-mass position of the corresponding fragment. Its radius
is chosen to be the maximum distance from the center at which the radial density
falls below \(\rho_{\mathrm{cut}}=10^{-4}~\mathrm{fm}^{-3}\). The
angular-momentum-projected kernels are calculated in the coordinate system
whose origin is fixed and co-moving with the center-of-mass position of the
fragment. In practice, all the single-particle wave functions after collision
in the whole space are subjected to the Gallilean boost to cancel the momentum
of the reaction product in $V$, as in Ref.~\cite{Sekizawa_2014}. The
gauge-angle integrals for PNP are evaluated using an equidistant discrete
Fourier mesh of \(95\) points. The Euler-angle integral for AMP is evaluated
using \(64\) equidistant points for \(\alpha\), a \(32\)-point Gauss--Legendre
quadrature in \(\cos\beta\), and \(128\) equidistant points over \(0\leq\gamma<4\pi\).
Angular-momentum quantum numbers up to \(J_{\max}=30\) are calculated. Rotated
single-particle wave functions at off-grid coordinates are evaluated using the
tensor-product cubic B-spline interpolation~\cite{Umar_1991}.

\section{Results and Discussion}\label{Sec:Results}

\subsection{Microscopic TDHF+PNP+AMP calculations for the multinucleon transfer reaction}

In this section, we present the results of the PNP+AMP analysis for
reaction products in the \(^{40}\mathrm{Ca}+{}^{124}\mathrm{Sn}\) reaction
at \(E_{\mathrm{c.m.}}=128.54~\mathrm{MeV}\). We find that the HF ground state
of the projectile, $^{40}$Ca, is of spherical shape, whereas that of the target,
$^{124}$Sn, is slightly deformed in an oblate shape with $\beta\simeq 0.11$,
consistent with the previous study \cite{Sekizawa_2013}. In TDHF calculations,
the symmetry axis of the oblately deformed $^{124}$Sn is set perpendicular
to the reaction plane at the initial time.

Ideally, when a deformed nucleus is involved in the reaction, one should
take an orientation average to evaluate cross sections that can directly be compared
with experimental data. In the present study, however, we disregard the orientation
dependence of the reaction and focus only on the representative orientation of
$^{124}$Sn mentioned above, since our main objective here is to demonstrate
the power of PNP+AMP and the richness of physics contained already in TDHF.
We point out here that the observed small deformation of $^{124}$Sn may be
an artifact of the HF approximation, because, in reality, it could be of
spherical shape in the presence of pairing correlations. This is an additional
reason why we consider that it is rational to postopone such a serious
investigatoin of the orientation dependence after taking into account
the pairing correlations.

The pairing correlations play, needless to say, an important role in both
nuclear structure and reaction dynamics~\cite{Ring_1980,Magierski_2017}. In recent years, it has become
possible to perform microscopic simulations taking full account of pairing
correlations based of the time-dependent Hartree--Fock--Bogoliubov (TDHFB) theory
or the time-dependent local density approximation (TDSLDA) in the context of
Kohn-Sham DFT for superfluid systems, using top-tier supercomputers~\cite{Bulgac_2016,Magierski_2022}.
However, as it makes calculations and interpretations far more complicated,
we neglect them, leaving an investigation of pairing correlations as a future task.
Nevertheless, we foresee and discuss possible effects of pairing correlations
on the results presented in the following sections.

As mentioned above, $^{124}\mathrm{Sn}$ is deformed in an oblate shape and
therefore contains $J\neq0\hbar$ components even at the initial state. In principle,
the $J=0\hbar$ component of $^{124}$Sn can be extracted with AMP for the initial state
and one may consider its time evolution, \textit{e.g.}, in a similar manner
as was done for PNP in Ref.~\cite{Scamps_2017}. However, this procedure
requires a number of separate TDHF calculations for all Euler angles and
the computational cost and complexity are exceedingly high. In the following
analysis, we thus focus exclusively on the PLF, which was a clean, doubly-magic
spherical $^{40}$Ca nucleus before collision.

\begin{figure}[t]
  \centering
  \includegraphics[width=\columnwidth]{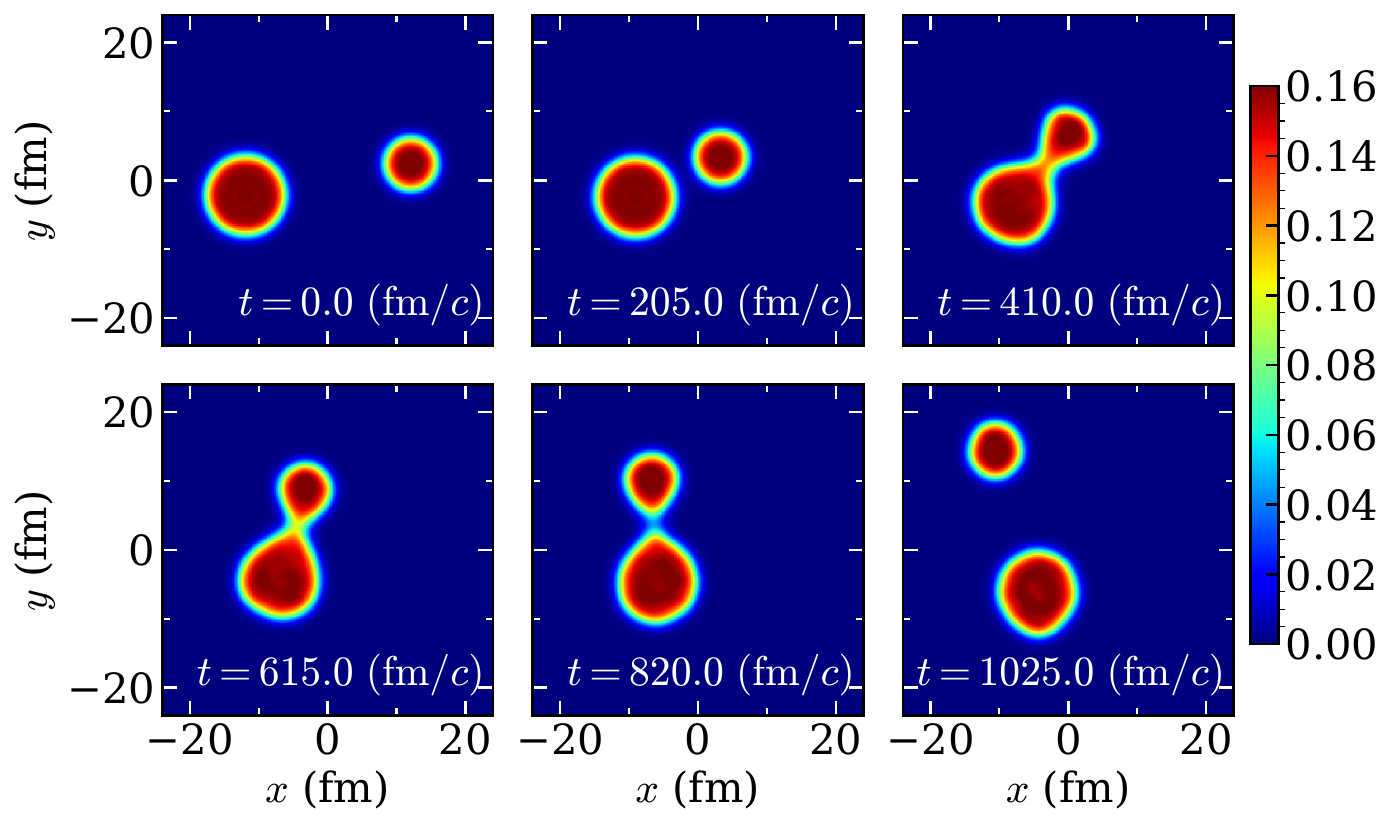}
  \caption{Snapshots of density distributions in the reaction plane at
  six representative times in the \(^{40}\mathrm{Ca}+{}^{124}\mathrm{Sn}\)
  reaction at \(E_{\mathrm{c.m.}}=128.54~\mathrm{MeV}\) and \(b=3.95~\mathrm{fm}\).
  }
  \label{fig:dynamics}
\end{figure}

To give an idea of the reaction dynamics under study, we show, in Fig.~\ref{fig:dynamics},
snapshots of cross sections of the density in the reaction plane for the
$^{40}$Ca+$^{124}$Sn reaction at $b=3.95$\,fm at six representative times.
The left-top panel ($t=0$\,fm/$c$) shows the initial configuration. As time
evolves, two nuclei approach each other ($t=205$\,fm/$c$) and form a neck
and exchange nucleons ($t=410$--$615$\,fm/$c$), and reseparate ($t=820$\,fm/$c$),
forming binary reaction products at the final time ($t=t_\text{f}=1025$\,fm/$c$).
This corresponds to the case of the smallest impact parameter that leads to
a binary reaction. Without the projection analysis, one can only integrate
the density around PLF and TLF to obtain the average number of transferred
nucleons. Even though the density looks simple, the whole system is described
by a single Slater determinant and significant information is contained in it.
In the following, we present the results of the PNP+AMP analysis and discuss
the underlying microscopic reaction mechanism.

\subsubsection{Zero-nucleon-transfer channel}\label{sec:0n0p}

\begin{figure}[t]
  \centering
  \includegraphics[width=0.48\textwidth]{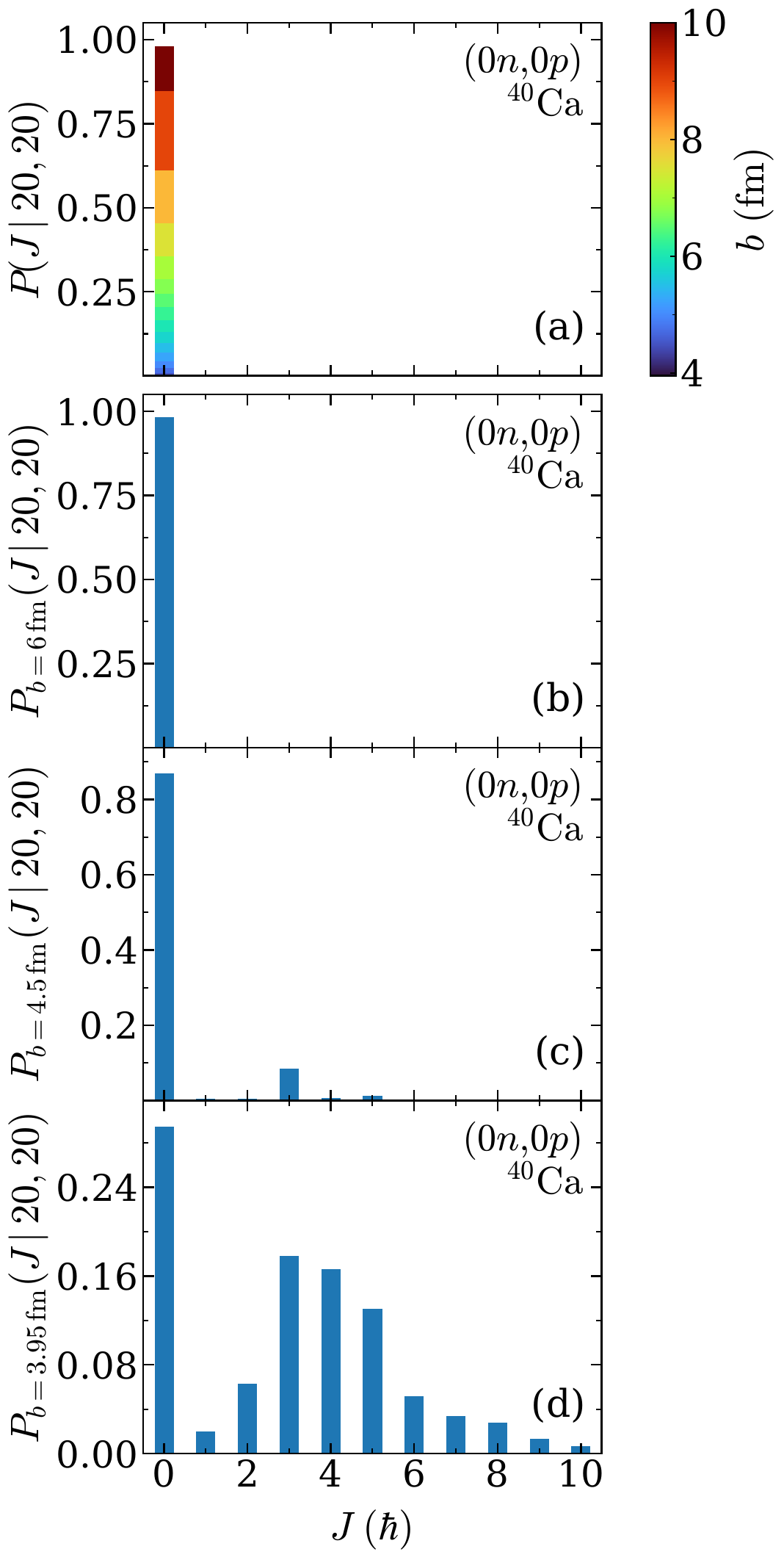}
  \caption{Angular-momentum distributions for the zero-nucleon-transfer channel, $^{40}\mathrm{Ca}$ [$(N,Z)=(20,20)$]. Panel (a) shows the impact-parameter-integrated distribution $P(J|20,20)$, while panels (b)--(d) show the corresponding impact-parameter-resolved distributions at $b=6$, $4.50$, and $3.95~\mathrm{fm}$, respectively.}
  \label{fig:zero_nucleon_angular}
\end{figure}

First of all, let us discuss the quasielastic channel without nucleon transfer,
$(0n,0p)$, producing $^{40}$Ca as a primay PLF. In this case, no angular momentum
is carried into or removed from the projectile and discussion becomes the simplest.
In Fig.~\ref{fig:zero_nucleon_angular}, we show the results of the PNP+AMP analysis
with repect to the PLF in the TDHF wave functions after collision. In all panels,
the horizontal axis corresponds to the total spin $J$. Figure~\ref{fig:zero_nucleon_angular}(a)
shows the impact-parameter-integrated ($b$-integrated) angular-momentum distribution,
$P(J|20,20)$, defined in Eq.~\ref{eq:conditional_j}. The contributions from different
impact parameters are shown as a stacked bar chart, where the corresponding $b$ value
is indicated in the color bar.

From Fig.~\ref{fig:zero_nucleon_angular}(a), we find that a pronounced peak
emerges at $J=0\hbar$ with a probability close to unity, suggesting that
the density distribution in the $(0n,0p)$ channel remains nearly spherical.
Larger contributions come from larger impact parameters, because not only
of the larger geometric contribution but the probability for $(0n,0p)$
decreases as other transfer channels open at smaller impact parameters.
Although the probability is close to unity, it is slightly less than 1,
indicating that $J>0$ excited states are also populated with particle-hole
excitations.

The population of such excited states can be identified by decomposing
the angualr-momentum distribution into contributions form each impact
parameter. Such an impact-parameter-identified ($b$-identified)
angular-momentum distributions are shown in Figs.~\ref{fig:zero_nucleon_angular}(b),
\ref{fig:zero_nucleon_angular}(c), and \ref{fig:zero_nucleon_angular}(d),
which show contributions from $b=6$\,fm, $4.5$\,fm, and $3.95$\,fm,
respectively. The $b$-identified angular-momentum distribution is
normalized such that $\sum_J P_b(J|N,Z)=1$. We note that the vertical
axis scale is different for Figs.~\ref{fig:zero_nucleon_angular}(c)
and \ref{fig:zero_nucleon_angular}(d).

When the impact parameter is large ($b\gtrsim 6~\mathrm{fm}$), the distribution
is strongly peaked at $J=0\hbar$, resembling the $b$-integrated distribution
shown in Fig.~\ref{fig:zero_nucleon_angular}(a). In contrast, in a pheripheral
collision ($b\simeq 4.5~\mathrm{fm}$), an additional peak appears at $J=3\hbar$,
as shown in Fig.~\ref{fig:zero_nucleon_angular}(c). This peak can be attributed
to the effect of octupole deformation induced by the neck formation. Experimentally,
the excited state of $^{40}\mathrm{Ca}$ has $J^\pi=3^{-}$~\cite{Kibedi_2002},
consistent with an RPA calculation using the octupole operator~\cite{Simenel_2013}.
In addition, Ref.~\cite{Simenel_2016} reported an increase in octupole deformation after
collision. In this way, with the PNP+AMP analysis, we can identify the nonzero
$J=3$ components associated with a collective excitation of $^{40}\mathrm{Ca}$.

As the impact parameter decreases further, other spin states are also populated
as shown in Fig.~\ref{fig:zero_nucleon_angular}(d). As can be seen from the figure,
there is a pronounced peak at $J=0\hbar$ and the second peak locates at $J=3\hbar$,
however, there are sizeable peaks at $J=4\hbar$ and $5\hbar$ as well, surrounded
with minor components from $J=1\hbar$ to $10\hbar$. The octupole excitation mode
alone cannot explain those spin states. There are two sources to populate such
a spin distribution at small impact parameters. One is the collective rotation,
where part of the relative orbital angular momentum is transferred to internal
motion that can induce rotation of reaction products. The other one is single-particle
excitations. For instance, in the case of one-particle-one-hole (1p-1h) excitation
from $d_{3/2}$, $s_{1/2}$, or $d_{5/2}$ state in the $sd$ shell to the initially
unoccipied $f_{7/2}$ state, addition of spins of a particle and a hole can generate
$J=2,3,\dots,5\hbar$, $J=3,4\hbar$, or $J=1,2,\dots,6\hbar$, respectively. Other
many-particle many-hole excitations are also possible in TDHF calculations.
The $b$-identified angular-momentum distribution shown in Fig.~\ref{fig:zero_nucleon_angular}(d)
can be regarded as a complex superposition of collective and single-particle
excitations in $^{40}$Ca, described by a single TDHF wave function after collision.

We note here that the results would not be affected even if we introduce
pairing correlations, because $^{40}$Ca is a doubly-magic nucleus, where
both neutrons and protons are in normal (unpaired) states. If an open-shell
nucleus were used for the projectile, particle-hole excitations will be surpressed
because of the pairing correlations. In such a case, the observed polulations
of the spin states other in Fig.~\ref{fig:zero_nucleon_angular}(d) (except
$J=0\hbar$ and $3\hbar$) could be altered, although it depends on the amount
of excitation energy if it is enough to break Cooper pairs.

\begin{figure*}[t]
  \centering
  \includegraphics[width=0.9\textwidth]{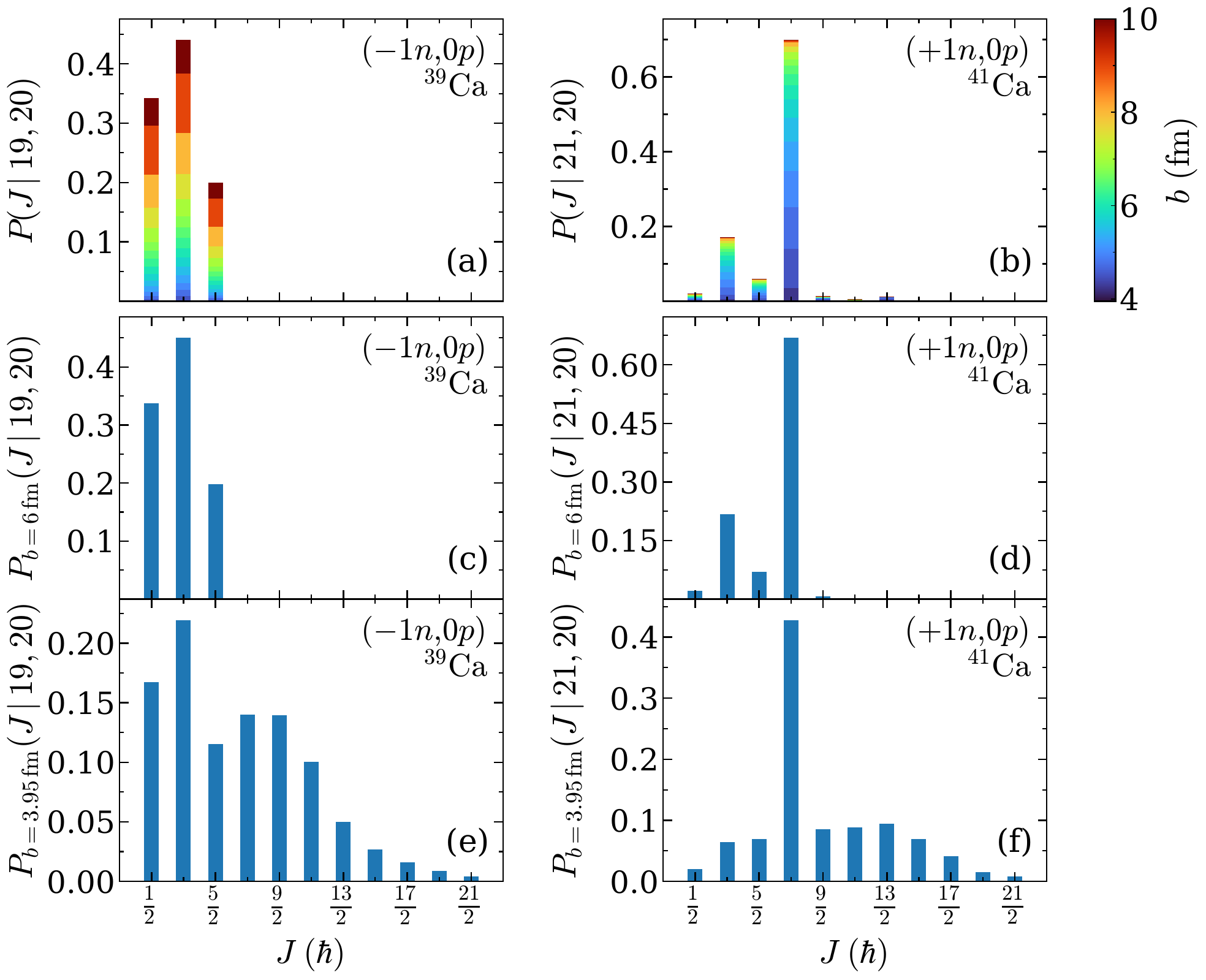}
  \caption{Angular-momentum distributions for the one-neutron-removal and one-neutron-addition channels of the PLF. Panels (a), (c), and (e) correspond to the one-neutron-removal channel, $^{39}\mathrm{Ca}$ [$(N,Z)=(19,20)$], whereas panels (b), (d), and (f) correspond to the one-neutron-addition channel, $^{41}\mathrm{Ca}$ [$(N,Z)=(21,20)$]. Panels (a) and (b) show the impact-parameter-integrated distributions $P(J|N,Z)$; panels (c) and (d), and panels (e) and (f), show the corresponding impact-parameter-resolved distributions at $b=6~\mathrm{fm}$ and $b=4~\mathrm{fm}$, respectively.}
  \label{fig:few_nucleon_angular}
\end{figure*}

\subsubsection{One-nucleon-transfer channels}

Next, let us move on to the analysis of one-nucleon transfer channels,
$(\pm1n,0p)$ and $(0n,\pm1p)$. In Fig.~\ref{fig:few_nucleon_angular}, we show
the angular momentum distributions, takeing the horizontal axis as the total
spin $J$ of the PLF. Left panels [Figs.~\ref{fig:few_nucleon_angular}(a),
\ref{fig:few_nucleon_angular}(c), and \ref{fig:few_nucleon_angular}(e)] show
the results obtained for the one-neutron removal channel ($-1n,0p$), while
right panels [Figs.~\ref{fig:few_nucleon_angular}(b), \ref{fig:few_nucleon_angular}(d),
and \ref{fig:few_nucleon_angular}(f)] show those for the one-neutron addtion
channel ($+1n,0p$). Since the results for the one-proton transfer channels
($0n,\pm1p$) are qualitatively similar to those for ($\pm1n,0$), we focus
here only on the latter case. [See, Supplemental Material \cite{SM}, for
the actual results for the ($0n,\pm1p$) case.] In Fig.~\ref{fig:few_nucleon_angular}(a),
we show the $b$-integrated angular-momentum distribution for the ($-1n,0p$) channel,
$P(J|19,20)$, as stacked bar chart, where contributions from different
impact parameters are indicated in the color bar, in the same way as
Fig.~\ref{fig:zero_nucleon_angular}(a). The $b$-identified angular-momentum
distributions for $b=6$\,fm and $3.95$\,fm are also shown in
Figs.~\ref{fig:zero_nucleon_angular}(c) and \ref{fig:zero_nucleon_angular}(e),
respectively.

From Fig.~\ref{fig:few_nucleon_angular}(a), we find that three pronounced peaks
emerge at $J=\frac{1}{2}\hbar$, $\frac{3}{2}\hbar$, and $\frac{5}{2}\hbar$.
Clearly, those peaks reflect the single-particle structure of $^{40}$Ca. Namely,
the first, second, and third heghest occupied single-particle states are
$d_{3/2}$, $s_{1/2}$, and $d_{5/2}$, respectively, and the order of the magnitude
of the populations matches with the order of these single-particle levels.
Thus, removal of one neutron with lower binding energy is more favarable
and the total angular momentum of the hole state determines the total spin
of the fragment. This naive picture holds when the impact parameter is large
($b\gtrsim 6$\,fm), as shown in Fig.~\ref{fig:few_nucleon_angular}(c). Again,
this intuitive interpretation holds when the impact parameter is large
($b\gtrsim 6$\,fm), as shown in Fig.~\ref{fig:few_nucleon_angular}(d).

A similar physical interpretation is possible in the case of the one-neutron
pickup channel ($+1n,0p$), shown in Fig.~\ref{fig:few_nucleon_angular}(b).
The relevant single-particle orbitals are ordered in energy as $f_{7/2}$,
$p_{3/2}$, $f_{5/2}$, $p_{1/2}$, and $g_{9/2}$, in ascending order, while
the dominant contributions to the angular-momentum distribtion, $P(J|21,20)$,
appear in the same order. Thus, when one neutron is added to $^{40}$Ca,
occupation of the $f_{7/2}$ orbital is energetically most favorable, and
it decreases with the change of states to $p_{3/2}$, $f_{5/2}$, $p_{1/2}$,
and $g_{9/2}$ states. The angular momentum of the additional one neutron
therefore determines the total angular momentum of the PLF.

We note that a closer look at Fig.~\ref{fig:few_nucleon_angular}(b) reveals
a tiny, but a non-zero population of $J=13/2\hbar$ in the $b$-integrated
angular-momentum distribution, which can not be explained solely by the
single-particle state of the additional neutron in the $pfg$ shell. It
indicates particle-hole excitations of the $^{40}$Ca core, similar to the
case of ($0n,0p$).

Such effects of particle-hole excitations can be seen in
Figs.~\ref{fig:few_nucleon_angular}(e) and \ref{fig:few_nucleon_angular}(f),
where the $b$-identified angular-momentum distributions for ($-1n,0p$) and
($+1n,0p$) channels are shown, respectively, for $b=3.95$\,fm. In the case
of the one-neutron removal channel ($-1n,0p$) in Fig.~\ref{fig:few_nucleon_angular}(e),
in addition to the $J=\frac{1}{2}\hbar$, $\frac{3}{2}\hbar$, and $\frac{5}{2}\hbar$
states, we find sizeable populations for $J=\frac{7}{2}\hbar$, $\frac{9}{2}\hbar$,
$\dots$, $\frac{21}{2}\hbar$. In the case of the one-neutron pickup channel
($+1n,0p$), on the other side, the population is still dominated by the
$J=\frac{7}{2}\hbar$ state, but fragmented states can also be found for
$J=\frac{1}{2}\hbar$, $\frac{3}{2}\hbar$, $\dots$, $\frac{21}{2}\hbar$.
It is interesting to recognize a qualitative difference between one-neutron
removal and pickup channels, where the former tends to excite more the
$^{40}$Ca core as compared to the latter case.

Similar to the ($0n,0p$) case discussed in Sec.~\ref{sec:0n0p}, because of
the doubly-magic character of $^{40}$Ca, the results would not be affected
even if we introduce pairing correlations, except for a change of probability
for one-nucleon transfer due to the presence of transfer of a correlated pair
of nucleons. Let us envision possible effects of pairing correlations
if an open-shell nucleus were used for the projectile. In the one-neutron
addition case, the spin populations associated with occupation of
single-particle states would not be changed qualitatively, since the
additional neutron is anyway unpaired. One may expect similar persistence
of single-particle picture also for the one-neutron removal case, although
the magnitude could be reduced as a Cooper pair must be broken to remove
a single neutron.

\subsubsection{Two-nucleon-transfer channels}

In this section, we discuss the case of two-nucleon-transfer channels,
($\pm2n,0p$) and ($0n,\pm2p$). Because, in the case of ($\pm1n,\pm1p$)
channels, the angular-momentum distribution is given by a simple addition
of those for neutrons ($\pm1n$) and protons $(\pm1p)$ and it can be trivially
explained with the Clebsch--Gordan coefficients.[See, Supplemental Material \cite{SM}, for the actual results for the ($\pm1n,\pm1p$) case.] Therefore, here we focus
only on the ($\pm2n,0p$) and ($0n,\pm2p$) channels.

\begin{figure}[t]
  \centering
  \includegraphics[width=\columnwidth]{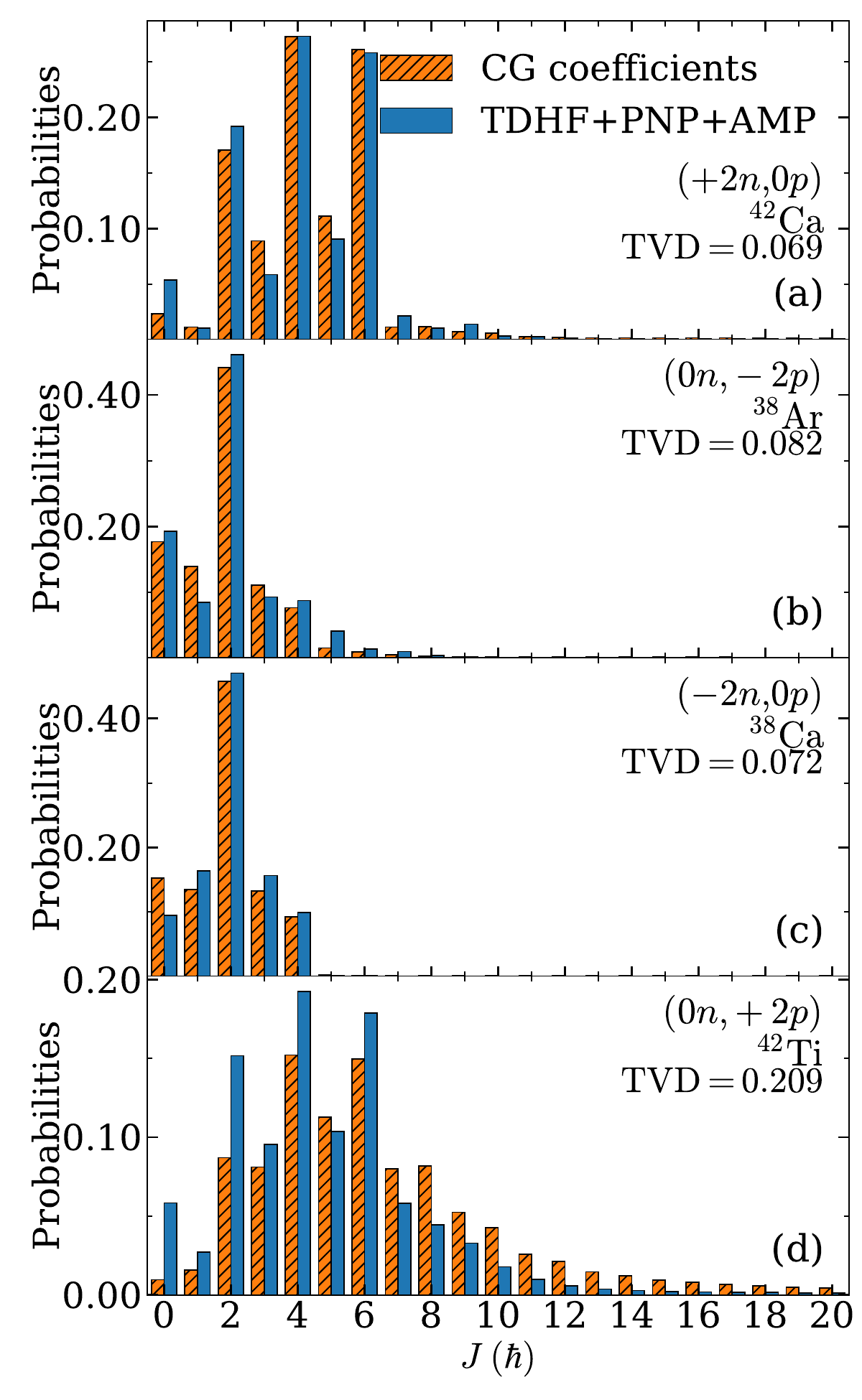}
  \caption{Comparison between the calculated angular-momentum distributions $P(J|N,Z)$ and the distributions $\tilde{P}_\mathrm{CG}(J|N,Z)$ reconstructed from the corresponding one-nucleon-transfer channels using the Clebsch--Gordan coupling model. The TVD values quantify the differences between the two distributions.
  }
  \label{fig:AMP_vs_CG}
\end{figure}

In Figs.~\ref{fig:AMP_vs_CG}(a), \ref{fig:AMP_vs_CG}(b), \ref{fig:AMP_vs_CG}(c),
and \ref{fig:AMP_vs_CG}(d), we show the $b$-integrated angular-momentum
distributions for ($+2n,0p$), ($0n,-2p$), ($-2n,0p$), and ($0n,+2p$) channels,
respectively. In all panels, the horizontal axis is the total spin of the PLF
in the corresponding channel. In each panel, the results obtained from the
PNP+AMP analysis for the TDHF wave functions after collision are shown by
blue filled bars. (The meaning of orange hatched bars will be explained below.)

As can be seen from Fig.~\ref{fig:AMP_vs_CG}, we find channel-dependent,
abundant spin distributions. For instance, in the case of the two-neutron
pickup channel ($+2n,0p$) shown in Fig.~\ref{fig:AMP_vs_CG}(a), there are
three pronounced peaks at $J=2\hbar$, $4\hbar$, and $6\hbar$. In addition,
there are peaks at $J=0\hbar$, $3\hbar$, and $5\hbar$ as well. Naively,
we can expect that the observed populations of those spin states are
originated from additions of contributions from single-particle states
of additional two nucleons at $f_{7/2}$, $p_{3/2}$, $f_{5/2}$, $p_{1/2}$,
or $g_{9/2}$, superposed over various combinations with proper weights.

To understand the observed spin distributions, we calculate spin distributions
assuming the single-particle picture and an isotropic distribution of the magnetic
substates. With these assumption, the angular-momentum distributions for
two-nucleon-transfer channels can be expressed as follows:
\begin{align}
    \tilde{P}_\mathrm{CG}(J|22,20) &= \sum_{j_1,j_2}\mathcal{P}_\mathrm{CG}(J|j_1,j_2)P(j_1|21,20)P(j_2|21,20), \\
    \tilde{P}_\mathrm{CG}(J|18,20) &= \sum_{j_1,j_2}\mathcal{P}_\mathrm{CG}(J|j_1,j_2)P(j_1|19,20)P(j_2|19,20), \\
    \tilde{P}_\mathrm{CG}(J|20,22) &= \sum_{j_1,j_2}\mathcal{P}_\mathrm{CG}(J|j_1,j_2)P(j_1|20,21)P(j_2|20,21), \\
    \tilde{P}_\mathrm{CG}(J|20,18) &= \sum_{j_1,j_2}\mathcal{P}_\mathrm{CG}(J|j_1,j_2)P(j_1|20,19)P(j_2|20,19),
\end{align}
where $\mathcal{P}_\mathrm{CG}$ is the Clebsch--Gordan coupling probabilities defined as
\begin{align}
    \mathcal{P}_\mathrm{CG}(J\mid j_1,j_2)
    &=
    \begin{cases}
        \displaystyle \frac{2J+1}{(2j_1+1)(2j_2+1)},
        & j_1\neq j_2,\\[4mm]
        \displaystyle \frac{2J+1}{\binom{2j_1+1}{2}},
        & 
         \begin{array}{l}
         j_1 = j_2,\\
         (-1)^{2j_1-J}=-1,
         \end{array}\\
        0,
        & \text{otherwise}.
    \end{cases}
\end{align}
These CG coupling probabilities account for the Pauli exclusion principle. 

The angular-momentum distributions obtined within the single-particle picture
$\tilde{P}_\mathrm{CG}(J|N,Z)$ are shown by orange hatched bars in Fig.~\ref{fig:AMP_vs_CG}
to compare with $P(J|N,Z)$ obtained from the PNP+AMP analysis (blue filled bars).
To quantify the difference between the two distributions, we define the total
variation distance (TVD) as
\begin{align}
    \mathrm{TVD}(N,Z)
    &= \frac{1}{2}\sum_J\left|\tilde{P}_\mathrm{CG}(J|N,Z)-P(J|N,Z)\right|.
\end{align}
The obtained TVD value is indicated in each panel of Fig.~\ref{fig:AMP_vs_CG}.

As shown in Fig.~\ref{fig:AMP_vs_CG}, we find a remarkable agreement between
$\tilde{P}_\mathrm{CG}(J|N,Z)$ and $P(J|N,Z)$. In all four panels, the reconstructed
CG distributions are quantitatively similar to the calculated distributions.
In particular, their peak positions agree well with each other, especially in
($\pm2n,0p$) and ($0n,-2p$) channels. Further, except for the $(0n,+2p)$
channel, the TVD values are small, ranging from $0.07$ to $0.08$, quantifying
the good agreement. The deviation observed in the ($0n,+2p$) channel can be
understood as follows. Since the main directions of nucleon transfers in
the $^{40}$Ca+$^{124}$Sn reaction are neutron pickup ($+xn$) and proton
stripping ($-xp$), the directions of the isospin equilibration. It means
that the two-proton pickup channel $(0n,+2p)$ is a minor channel, which
occurs only at small impact parameters, with a very small probability
($\sim 10^{-4}$). At such small impact parameters, as discussed in previous
sections, $(n+2)$p-$n$h excitations are induced, resulting in the observed
deviation with a somewhat larger TVD value. Nevertheless, these results indicate
that the single-particle picture holds well, not only for the one-nucleon-transfer
channels, but also for the two-nucleon-transfer channels.

One should note, however, that the overseved agreement between the microscopic
PNP+AMP analysis and the pure single-particle picture is partly due to the
neglected pairing correlations. Namely, if the pairing correlations are
intdocuded, transfers of correlated Cooper pairs can contribute to the
two-nucleon transfer channels, in addition to the sequencial transfers
of uncollelated nucleons. Since the correlated nucleons predominantly
form $J=0\hbar$ Cooper pairs, the angular-momentum distributions would
deviate from the single-particle estimate. Intriguingly, in other words,
a close comparison between precisely-measured spin distributions of
reaction products and the single-particle estimate would provide a hint
on the effects of pairing correlations in two-nucleon-transfer processes.

\subsubsection{Many-nucleon-transfer channels}

In this section, we discuss evolusion of spin distributions in reaction products
for channels with more than two nucleons.

\begin{figure}[t]
  \centering
  \includegraphics[width=0.48\textwidth]{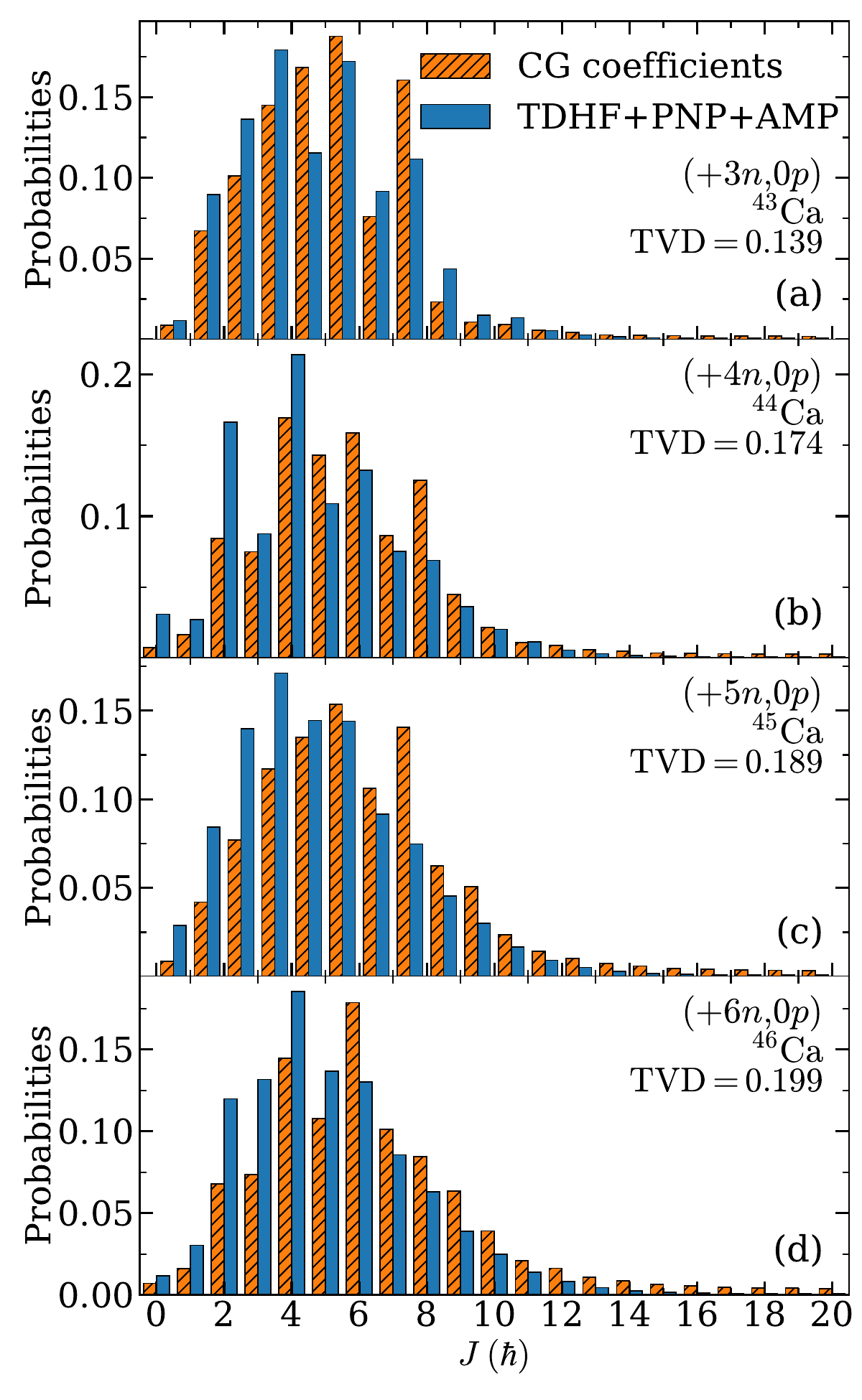}
  \caption{Comparison between the calculated angular-momentum distributions $P(J|N,Z)$ and the CG-reconstructed distributions $\tilde{P}_{\mathrm{CG}}(J|N,Z)$ for the neutron-transfer channels (a) $^{43}\mathrm{Ca}$ $(+3n,0p)$, (b) $^{44}\mathrm{Ca}$ $(+4n,0p)$, (c) $^{45}\mathrm{Ca}$ $(+5n,0p)$, and (d) $^{46}\mathrm{Ca}$ $(+6n,0p)$. The CG-reconstructed distributions are obtained using the Clebsch--Gordan coupling model, and the TVD values quantify the differences between the two distributions in each panel.}
  \label{fig:AMP_vs_CG_345n}
\end{figure}

Following the procedure of Ref.~\cite{Pain_2018,Poirier_2021}, we extend the CG coupling model based on
the single-particle picture, which was introduced in the previous section, for
channels involving an arbitrary number of transferred nucleons.
In Figs.~\ref{fig:AMP_vs_CG_345n}(a), \ref{fig:AMP_vs_CG_345n}(b),
\ref{fig:AMP_vs_CG_345n}(c), and \ref{fig:AMP_vs_CG_345n}(d), we compare the
CG-reconstructed angular-momentum distributions (orange hatched bars) and
those obtained from the microscopic PNP+AMP analysis (blue filled bars) for
the $(+3n,0p)$, $(+4n,0p)$, $(+5n,0p)$, and $(+6n,0p)$ channels, respectively.
Since the probability decreases rapidly with increasing the number of tranasferred
protons, here we focus on the pure neutron-transfer channels without proton transfer.
From the results, we find that the peak positions of the CG-reconstructed
distributions deviate from those of the calculated distributions, in contrast
to the successful agreement observed in Figs.~\ref{fig:AMP_vs_CG}(a)--(c).
It is also quantified by the TVD value, which increases with the number of
transferred neutrons. The observed deviation suggests that collective effects
and multiparticle--multihole excitations become important in many-nucleon-transfer
channels.

\begin{figure}[t]
  \centering
  \includegraphics[width=0.48\textwidth]{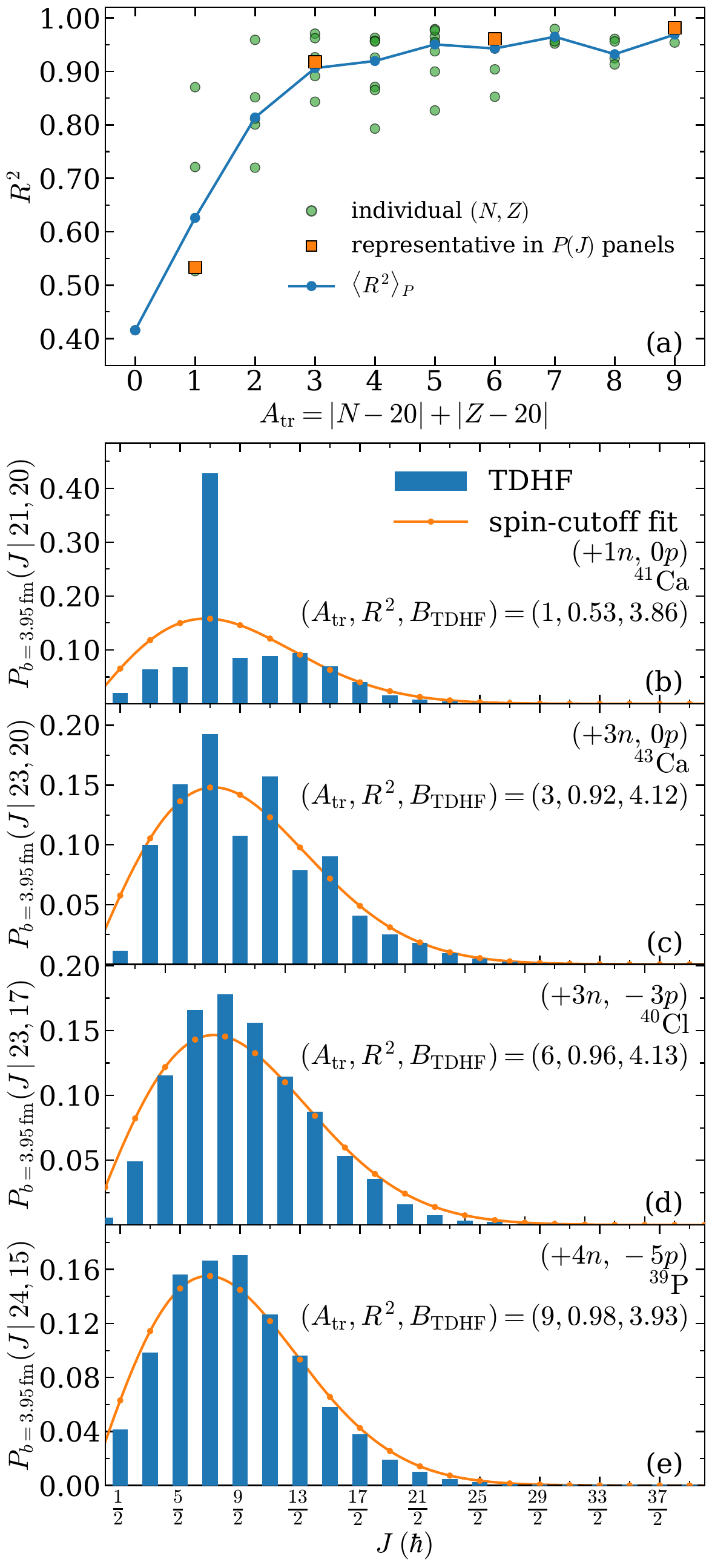}\vspace{-2mm}
  \caption{Spin-cutoff fits to the TDHF angular-momentum distributions. (a) $R^2$ versus $A_{\mathrm{tr}}=|N-20|+|Z-20|$ for channels with $P(N,Z)\geq 10^{-4}$ (green circles), their $P(N,Z)$-weighted mean (blue line), and channels selected for panels (b)--(e) (orange squares). (b)--(e) TDHF distributions (blue) and fits (orange) for $A_{\mathrm{tr}}=1$, 3, 6, and 9.}\vspace{-2mm}
  \label{fig:many_nucleon_angular}
\end{figure}

\begin{figure}[t]
  \centering
  \includegraphics[width=0.48\textwidth]{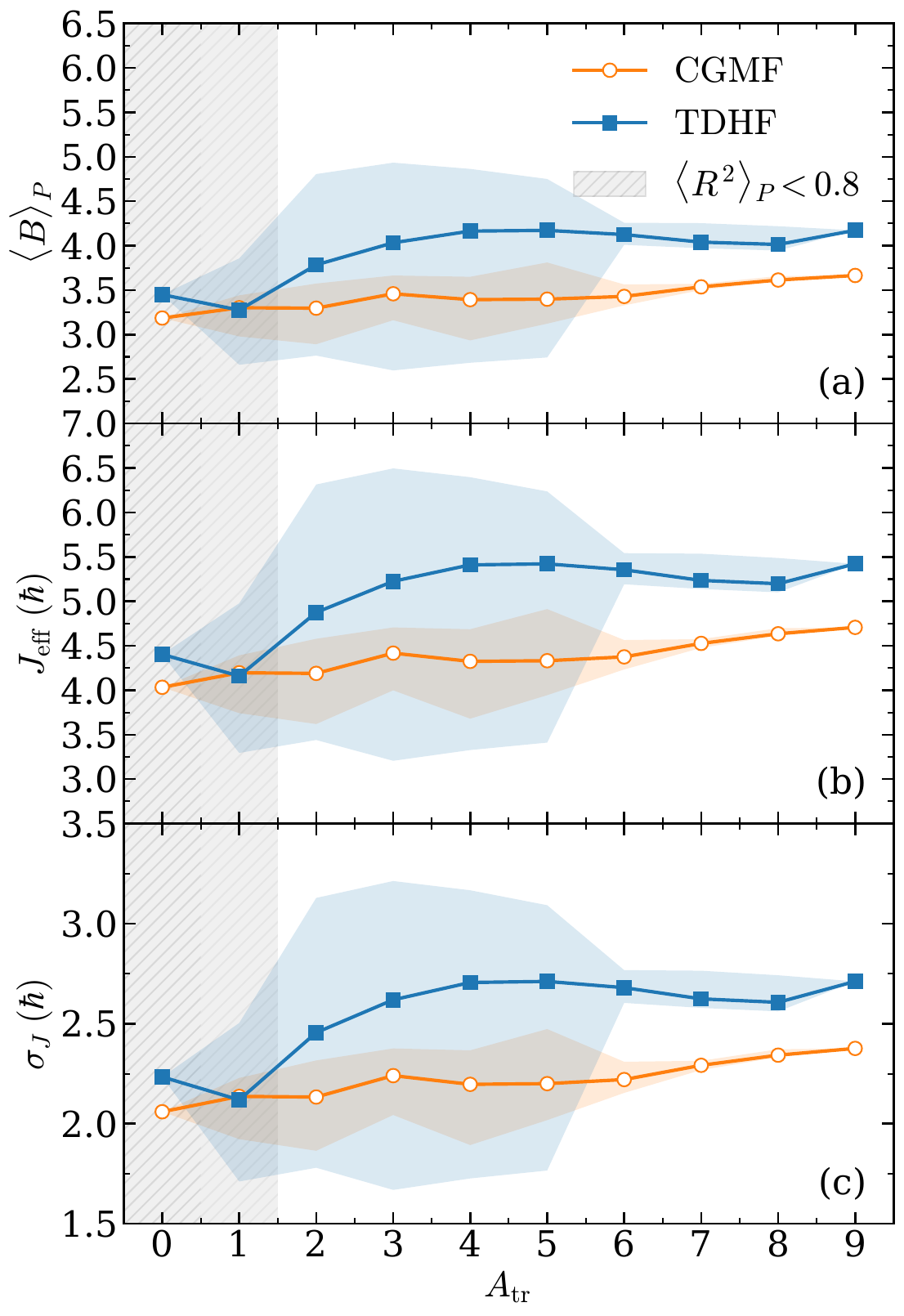}
  \caption{Comparison of the spin-cutoff parameters obtained from the TDHF distributions and the CGMF prescription as a function of $A_{\mathrm{tr}}$. Symbols show the $P(N,Z)$-weighted means, and the shaded bands indicate the ranges across the contributing $(N,Z)$ channels.}
  \label{fig:B_TDHF_vs_B_CGMF}
\end{figure}

The observed angular-momentum distributions for many-nucleon-transfer channels
remind us of spin-distributions in fission fragments. For example, microscopic
studies of fission fragments reported angular-momentum distributions with shapes
resembling empirical spin-cutoff distributions~\cite{Marevic_2026,Zhang_2025}.
In the statistical-model code CGMF~\cite{Talou_2021}, the initial fragment
angular-momentum distribution is assumed to have the spin-cutoff form,
\begin{align}
    P_{\mathrm{CGMF}}(J|N,Z) \propto (2J+1)\exp\left[-\frac{J(J+1)}{2B^2(Z,A,T)}\right],
    \label{eq:spin-cutoff}
\end{align}
where $B(Z,A,T)$ is the spin-cutoff parameter, defined in CGMF by $B^2(Z,A,T)
=\alpha\mathcal{I}_0(A,Z)T/\hbar^2$. Here, $\mathcal{I}_0(A,Z)$ is the ground-state
moment of inertia of the corresponding nucleus, $T$ is the fragment temperature
determined from its excitation energy through the level-density model, and
$\alpha$ is a global adjustable parameter which is adjusted to reproduce
prompt-fission $\gamma$-ray data.

With this spin-cutoff distribution, we carried out a fitting to the
angular-momentum distirubtions of the PNP+AMP analysis, and the results
are shown in Fig.~\ref{fig:many_nucleon_angular}.
In Figs.~\ref{fig:many_nucleon_angular}(b)--(e), we show the calculated
angular-momentum distributions (blue filled bars) and their spin-cutoff fits
(orange lines) for representative channels with different total number of
transferred nucleons, $A_{\mathrm{tr}}=1$, $3$, $6$, and $9$, respectively.
Because collisions at smaller impact parameters are expected to drive the
system closer to the statistical equilibrium, we focus here on the distributions
obtained at $b=3.95~\mathrm{fm}$. From the results, we find that the fit quality
is improved with increasing $A_{\mathrm{tr}}$, as indicated by the larger
coefficient of determination, the $R^2$ values. The whole results are
summarized in Fig.~\ref{fig:many_nucleon_angular}(a). 
Figure~\ref{fig:many_nucleon_angular}(a) shows the $R^2$ values for individual
$(N,Z)$ channels satisfying $P(N,Z)\geq 10^{-4}$, together with the $P(N,Z)$-weighted
mean for each $A_{\mathrm{tr}}$. Both the selected channels and the $P(N,Z)$-weighted
means show better agreement with the spin-cutoff form as $A_{\mathrm{tr}}$ increases.
Qualitatively, this trend suggests that the angular-momentum distributions approach
the statistical-equilibrium form as more nucleons are transferred. In the single-particle
picture, this behavior may reflect the increasing number of angular-momentum-coupling
configurations, associated with the increasing complexity of many-particle--many-hole
excitations.

As we found a qualitative agreement with the spin-cutoff distribution, as a next step,
we quantitatively compare the microscopic distributions with the empirical spin-cutoff
prescription. Determining $B_{\mathrm{CGMF}}$ requires the fragment excitation energy $E^*$. Following Refs.~\cite{Sekizawa_2017,Wu_2019,Jiang_2018,Jiang_2020} and using
the AME2020 mass data~\cite{Huang_2021,Wang_2021}, we estimate $E^*$ and set the
global adjustable parameter $\alpha$ to unity. It is to mention here that we have
also tested the excitation energy estimation including a Coulomb correction as
discussed in Ref.~\cite{sekizawa_2020}, but the correction turned out to be
negligible in the present case. The latter correction will be important when
channels involving transfer of many protons are investigated.
In Fig.~\ref{fig:B_TDHF_vs_B_CGMF}(a), we compare
the $P(N,Z)$-weighted mean values of $B_{\mathrm{TDHF}}$ and $B_{\mathrm{CGMF}}$,
while the shaded bands indicate the corresponding ranges over the contributing
channels. The CGMF averages are likewise weighted by the TDHF transfer probabilities
$P(N,Z)$. [See, Supplemental Material \cite{SM}, for the datas.] At $A_{\mathrm{tr}}=1$,
$\expval{B_{\mathrm{TDHF}}}_P$ and $\expval{B_{\mathrm{CGMF}}}_P$ appear comparable.
However, the TDHF value is not reliable because the weighted mean $\expval{R^2}_P$
is small. A clear discrepancy emerges at larger $A_{\mathrm{tr}}$. Namely, at
$A_{\mathrm{tr}}=9$, for instance, the relative difference is $13.8\%$. For the
spin-cutoff distribution in Eq.~\eqref{eq:spin-cutoff}, the effective angular
momentum $J_\mathrm{eff}$~\cite{Marevic_2026,Bulgac_2021,Marevic_2021} and the
distribution width $\sigma_J$ are defined as
\begin{align}
    J_\mathrm{eff} &= \frac{\sqrt{1+8B^2}-1}{2},\\
    \sigma_J &=\sqrt{2B^2-\expval{J}-\expval{J}^2},
\end{align}
where
\begin{align}
    J_\mathrm{eff}(J_\mathrm{eff}+1) \equiv \expval{J(J+1)} = 2B^2,\\[2mm]
    \expval{J} = B\sqrt{\frac{\pi}{2}}\exp{\frac{1}{8B^2}}\mathrm{erfc}\left\{ \frac{1}{2\sqrt{2}B} \right\}.
\end{align}
In Fig.~\ref{fig:B_TDHF_vs_B_CGMF}(c) and \ref{fig:B_TDHF_vs_B_CGMF}(d),
we show $J_\mathrm{eff}$ and $\sigma_J$, respectively. At $A_{\mathrm{tr}}=9$,
the TDHF values of $J_\mathrm{eff}$ and $\sigma_J$ exceed the corresponding
CGMF values by approximately $0.7~\hbar$ and $0.3~\hbar$, respectively.
This indicates that the microscopic distributions have larger effective
angular momenta and broader widths as compared to those predicted by the
empirical spin-cutoff prescription.

Summarizing the analsysis for the many-nucleon-transfer channels,
we have demonstrated that the angular-momentum distributions obtained
from the TDHF wave function after collision with the PNP+AMP method evolve
towards the statistical equilibrium, resembling the spin-cutoff distributions
assumed in statistical models. Moreover, the angular-momentum distributions
in TDHF are even wider than that of the CGMF estimate. One may find the results
surprising, since it has been well known that the TDHF fails severely to reproduce
the width of distributions of observables due to the single-Slater approximation
and one had to use extended approaches such as time-dependent random phase approximation
(TDRPA) to cure the drawback. The reason why we could obtain the wide angular-momentum
distributions from the TDHF wave function is that those distributions are originated
mainly from complex couplings of single-particle degrees of freedom. Because there are
lots of single-particle states with a huge number of possible excited configurations
even within a single mean-field potential, the TDHF theory could describe abundant
fragment spin distributions in multinucleon transfer reactions.

We mention here that the results for the many-nucleon-transfer channels would not
be affected much by the neglected pairing correlations, since those channels accompany,
in general, substantial internal excitations that would break Cooper pairs.

\subsubsection{Time dependence of the angular-momentum distribution}\label{sec:td-dist-MNT}

\begin{figure}[t]
  \centering
  \includegraphics[width=0.48\textwidth]{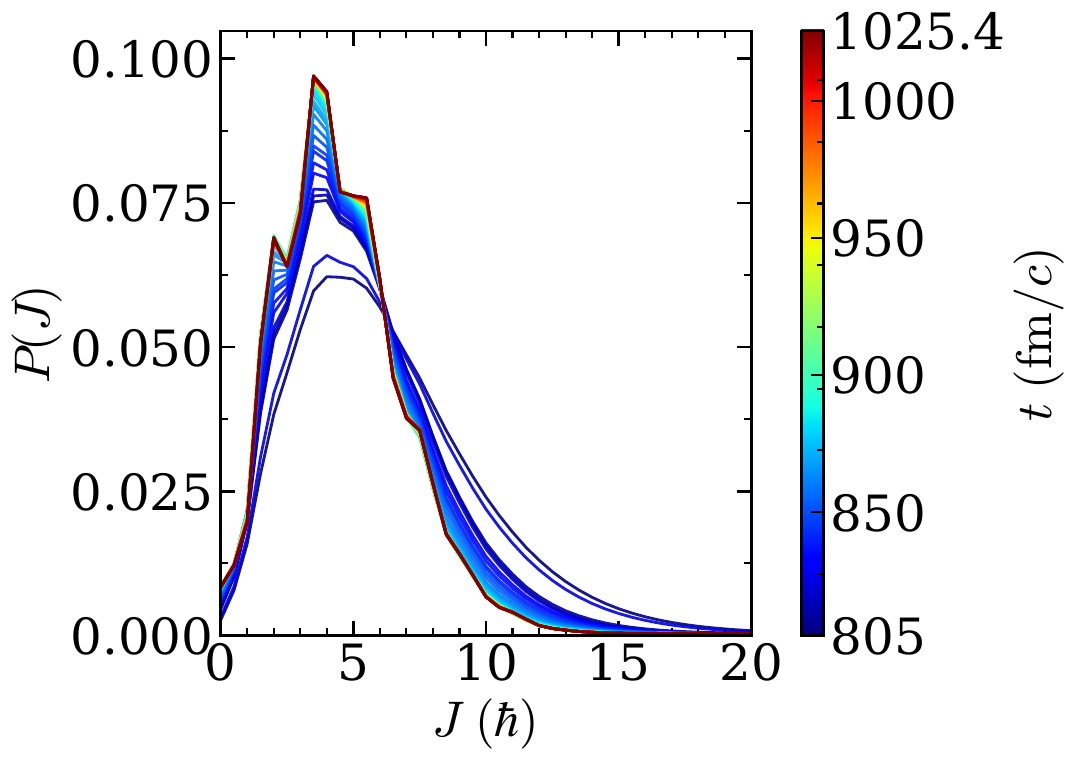}
  \caption{Channel-summed angular-momentum distributions of the PLF at intervals of $5~\mathrm{fm}/c$ from $t=805$ to $1025.4~\mathrm{fm}/c$.}
  \label{fig:P(J)_time}
\end{figure}

Nevertheless our main resluts have been given in the previous sections, here we
report an additional analysis that may call certain attention. Namely, in this section,
we examine time dependence of the angular-momentum distribution after collision.
To this end, we perform PNP+AMP for $b=3.95$\,fm at different instances after collision.
In Fig.~\ref{fig:P(J)_time}, we show the total angular-momentum distribution, $P(J)$,
calculated from $t=805$ to $1025.4~\mathrm{fm}/c$ with $5~\mathrm{fm}/c$ interval.
Before $t=895~\mathrm{fm}/c$, no point between the fragments satisfies
$\rho<\rho_{\mathrm{cut}}=10^{-4}$, and the radius of $V$ is therefore determined
from the location of the minimum density. Consequently, the calculated distribution
changes substantially with time during this interval. Before complete separation,
the distribution resembles the empirical spin-cutoff form. Then, after separation,
characteristic peaks associated with the shell structure emerge. In this way,
we can witness emergence of the individual character such as shell structure
as a fragmented nucleus.

 \begin{figure}[t]
  \centering
  \includegraphics[width=0.4\textwidth]{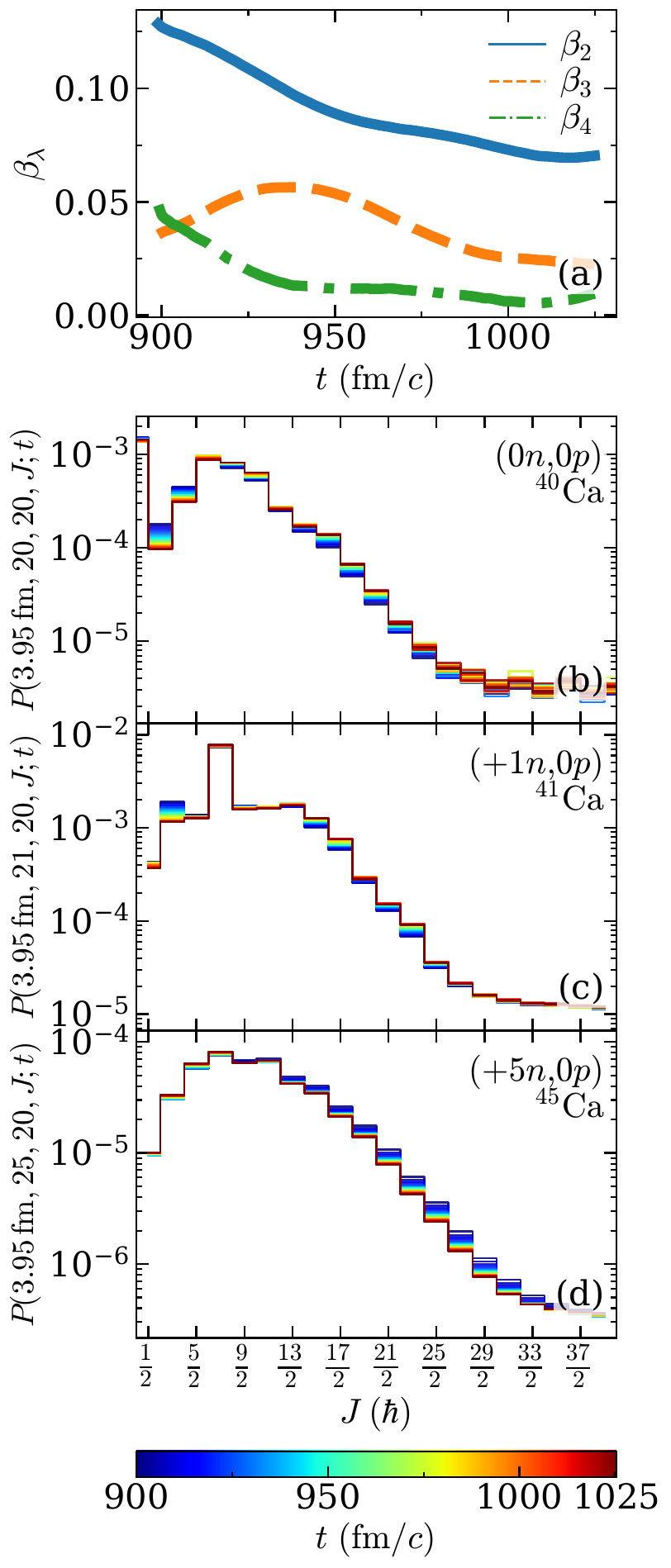}
  \caption{Time evolution of the PLF deformation parameters $\beta_2$, $\beta_3$, and $\beta_4$ and the angular-momentum distributions for selected $(N,Z)$ channels at $b=3.95~\mathrm{fm}$. The distributions are evaluated at intervals of $5~\mathrm{fm}/c$ after PLF--TLF separation.}
  \label{fig:beta234_amp}
\end{figure}

To get deeper insight into the fragment spin evolution, we show, in Fig.~\ref{fig:beta234_amp},
the $b$-identified angular-momentum distributions for $b=3.95$\,fm in three representative
reaction cahnels, ($0n,0p$) [panel (b)], ($+1n,0p$) [panel (c)], and ($+5n,0p$) [panel (d)].
The angular-momentum distributions are calculated every $5~\mathrm{fm}/c$ after separation
and the corresponding times are represented by line colors. Although small variations are present,
the overall shape of each distribution remains roughly unchanged. From Fig.~\ref{fig:beta234_amp}(b),
we find that both the $J=0\hbar$ component, reflecting the sphericity of ${}^{40}\mathrm{Ca}$,
and the $J=3\hbar$ component, associated with the dominant collective excitation induced by
the neck formation between colliding nuclei, remain nearly constant. Also, in the case of
one-neutron pickup channel shown in Fig.~\ref{fig:beta234_amp}(c), the largest peak at
$J=7/2\hbar$, which originates from the occupation of $f_{7/2}$ single-particle state,
is unchanged. In the case of many-neutron transfer shown in Fig.~\ref{fig:beta234_amp}(d),
the spin-cutoff-like distribution is maintained throughout the time evolution, while
a slightly decreasing trend of high-spin components can be observed. The observed weak
time dependence of spin distributions may indicate the conservation of the fragment angular
momentum in its subsequent shape evolution. Actually, time evolution of deformation parameters,
$\beta_2$, $\beta_3$, and $\beta_4$, of a PLF after reseparation is served as supplemental
information in Fig.~\ref{fig:beta234_amp}(a) for the $b=3.95~\mathrm{fm}$ case. From
Fig.~\ref{fig:beta234_amp}(a), we find a decreasing trend of all deformation parameters
as a function of time, which may partly be related to the slightly decreasing trend of
high-spin components shown in Fig.~\ref{fig:beta234_amp}(c).

To conclude, we consider that the observed weak time dependence of fragment spin
distributions confirms the robustness of the PNP+AMP results shown in the preceeding sections.

\subsection{Microscopic TDHF+PNP+AMP calculations for induced fission of ${}^{240}\mathrm{Pu}$}

So far, we have focused on the angular-momentum distributions of PLFs generated through
multinucleon transfer processes in the $^{40}$Ca+$^{124}$Sn reaction. Within the same
framework, we can also analyze spin disrtibutions in fission products. In this section,
we report exploratory results of the PNP+AMP analysis applied for nuclei generated
through induced fission of $^{240}$Pu.

\begin{figure}[t]
  \centering
  \includegraphics[width=0.48\textwidth]{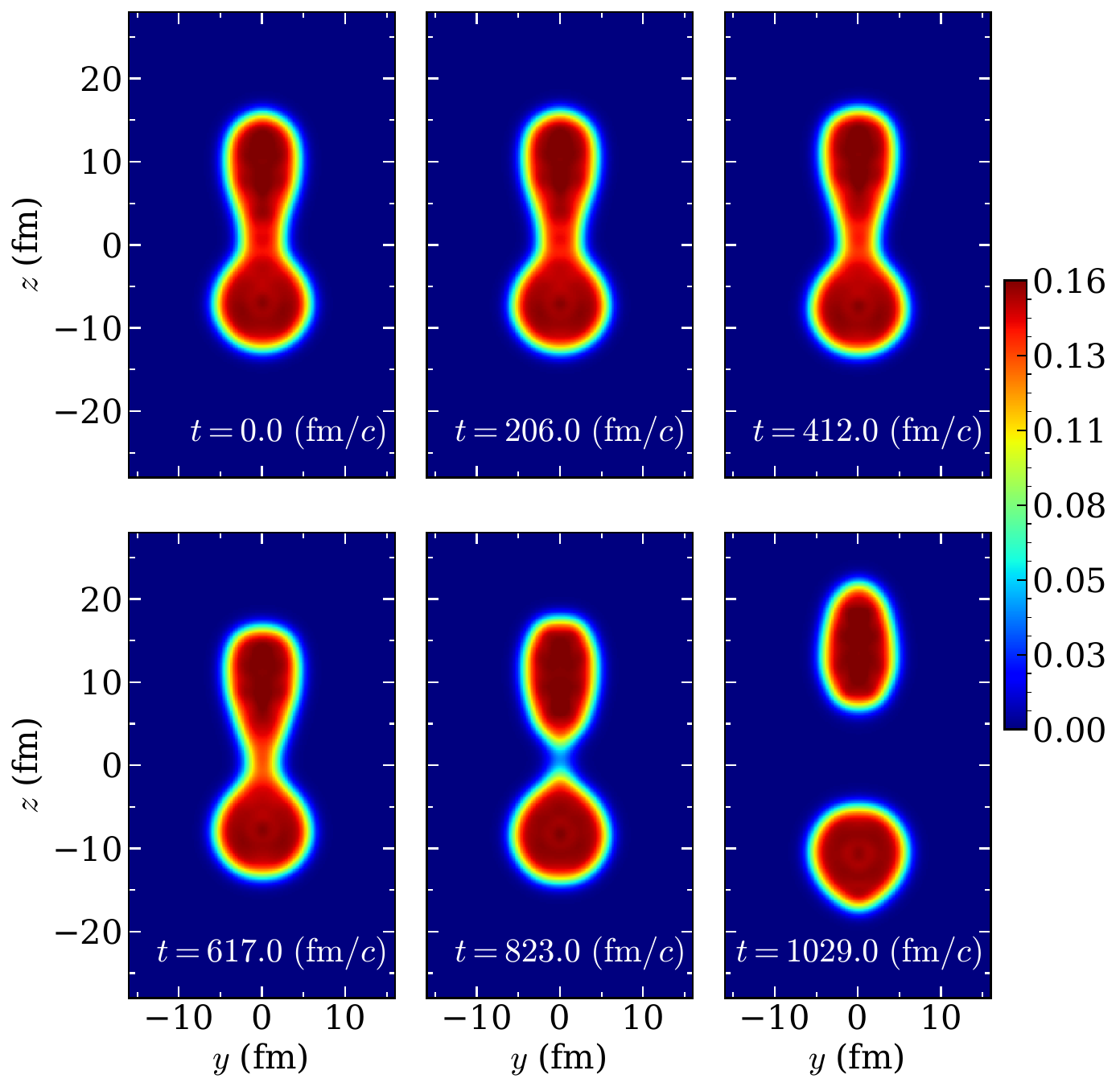}
  \caption{Snapshots of density distributions in the plane which parallel to intrinsic axis for quadrupole deformation at six representative times in the induced fission of ${}^{240}$Pu.}
  \label{fig:ind_fission}
\end{figure}

In Fig.~\ref{fig:ind_fission}, we show snapshots of the density during the induced fission
of $^{240}\mathrm{Pu}$. We employ the same EDF as in the reaction calculations. The initial
state is obtained from a constrained Hartree--Fock (CHF) calculation at $\beta_2=3$ imposing
the axial symmetry. Although this highly elongated initial state lies well beyond the fission
barrier, it is adopted to reduce the computational cost and to provide a sufficiently long
post-scission evolution for examining the time dependence of the distributions. We thus
restrict the present discussion to a qualitative analysis of the shapes and time dependence
of the angular-momentum distributions. At the final time of the simulation, $t=1029.4~\mathrm{fm}/c$,
the average neutron and proton numbers $(\expval{N},\expval{Z})$ of the lighter (upper side)
and heavier (lower side) fragments are calculated to be around $(\expval{N}_\text{L},\expval{Z}_\text{L})
=(63,41)$ and $(\expval{N}_\text{L},\expval{Z}_\text{L})=(83,53)$, respectively. In the PNP+AMP analysis
for fission fragments, we set the upper limit of spin distributions to $J=40\hbar$.

\begin{figure*}[t]
  \centering
  \includegraphics[width=0.8\textwidth]{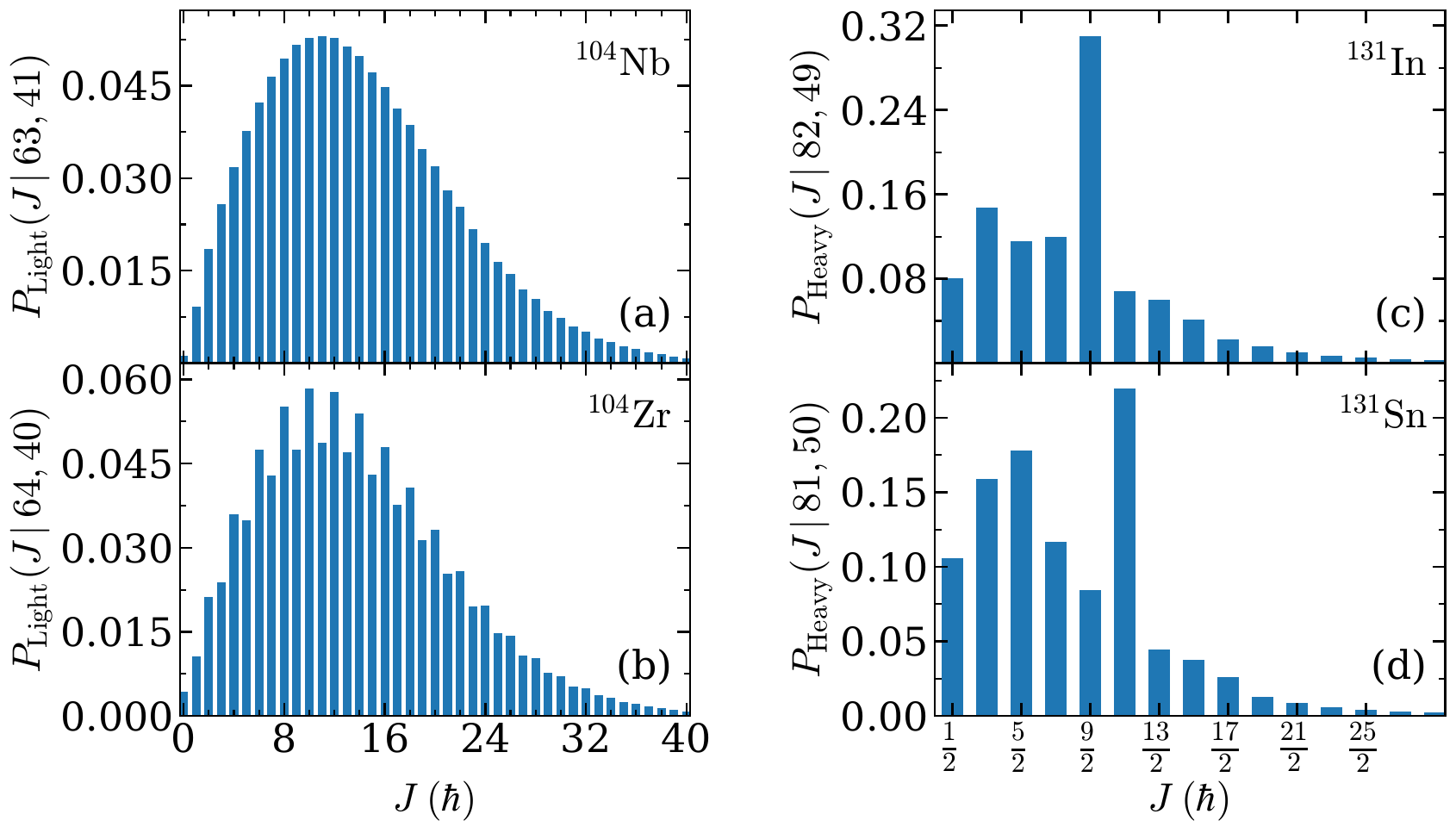}
  \caption{Angular-momentum distributions of selected fission-fragment channels at the final TDHF time. Panels (a) and (b) show the light-fragment distributions for $(N,Z)=(63,41)$ and $(64,40)$, respectively, while panels (c) and (d) show the heavy-fragment distributions for $(N,Z)=(82,49)$ and $(81,50)$, respectively.}
  \label{fig:fission_P(J)}
\end{figure*}

The results of the PNP+AMP calculations for lighter (heavier) fragments are
shown in left (right) panels of Fig.~\ref{fig:fission_P(J)}. In Figs.~\ref{fig:fission_P(J)}(a),
\ref{fig:fission_P(J)}(b), \ref{fig:fission_P(J)}(c), and \ref{fig:fission_P(J)}(d),
the angular-momentum distributions of specific fission fragments, $^{104}$Nb,
$^{104}$Zr, $^{131}$In, and $^{131}$Sn, are shown, respectively, as representative
examples. We note that the PNP+AMP calculations for the lighter and heavier fragments
were performed separately, taking the analyzing volume $V$ around each of those fragments.

First, by looking at Figs.~\ref{fig:fission_P(J)}(a) and \ref{fig:fission_P(J)}(b),
we find that the angular-momentum distribution of the lighter fragment exhibits
a typical spin-cutoff-like shape, consistent with previous microscopic
studies~\cite{Marevic_2026,Scamps_2026}. On top of the spin-cutoff-like distribution,
we find some odd--even staggering in Fig.~\ref{fig:fission_P(J)}(b), indicating
rotational-band-like distributions at even values of $J$. We consider that it
reflects the large fragment deformation of $\beta_2^\mathrm{Light}\sim0.7$.

In stark contrast, the average neutron and proton numbers of the heavier fragment
lie near the magic numbers, $N=82$ and $Z=50$, and its density distribution is
nearly of spherical shape with $(\beta_2^\mathrm{Light},\beta_3^\mathrm{Light},\beta_4^\mathrm{Light})\sim(0.05,0.06,0.03)$. Because of this small deformation,
the shell effects in the angular-momentum distributions are visible, as in the
multinucleon transfer reaction. In Fig.~\ref{fig:fission_P(J)}(c) and \ref{fig:fission_P(J)}(d),
we show the angular-momentum distributions of the $(N_\text{H},Z_\text{H})=(82,49)$
and $(81,50)$ channels at the final time, respectively. Clearly, both distributions
do not show the spin-cutoff-like pattern, unlike the lighter fragments. For the
SLy5 EDF, the single-particle orbitals in the shell between $28$ and $50$ are:
$1g_{9/2}$, $2p_{1/2}$, $2p_{3/2}$, and $1f_{5/2}$ in descending order in energy
(see, the Supplemental Material \cite{SM}, for details of single-particle structure
in the HF ground states). From Fig.~\ref{fig:fission_P(J)}(c), we find clear peaks
at $J=\frac{9}{2}\hbar$ and $\frac{3}{2}\hbar$, together with relatively large
probabilities at $J=\frac{1}{2}\hbar$, $\frac{5}{2}\hbar$, and $\frac{7}{2}\hbar$.
Similarly, the orbitals in between shells of $50$ and $82$ are: $1h_{11/2}$,
$2d_{3/2}$, $3s_{1/2}$, $1g_{7/2}$, and $2d_{5/2}$ in descending order in energy.
From Fig.~\ref{fig:fission_P(J)}(d), we find a prominent peak at $J=\frac{11}{2}\hbar$
and relatively large probabilities at $J=\frac{1}{2}\hbar$, $\frac{3}{2}\hbar$,
$\frac{5}{2}\hbar$, and $\frac{7}{2}\hbar$. Similar features also appear in the
$(N_\text{H},Z_\text{H})=(82,51)$ and $(83,50)$ channels. In this way, part of
peaks can be explained within a simple single-particle picture in induced fission,
when shell effects remain evident, albeit less prominently than in the case of
the multinucleon transfer reactions.

\begin{figure}[bt]
  \centering
  \includegraphics[width=0.48\textwidth]{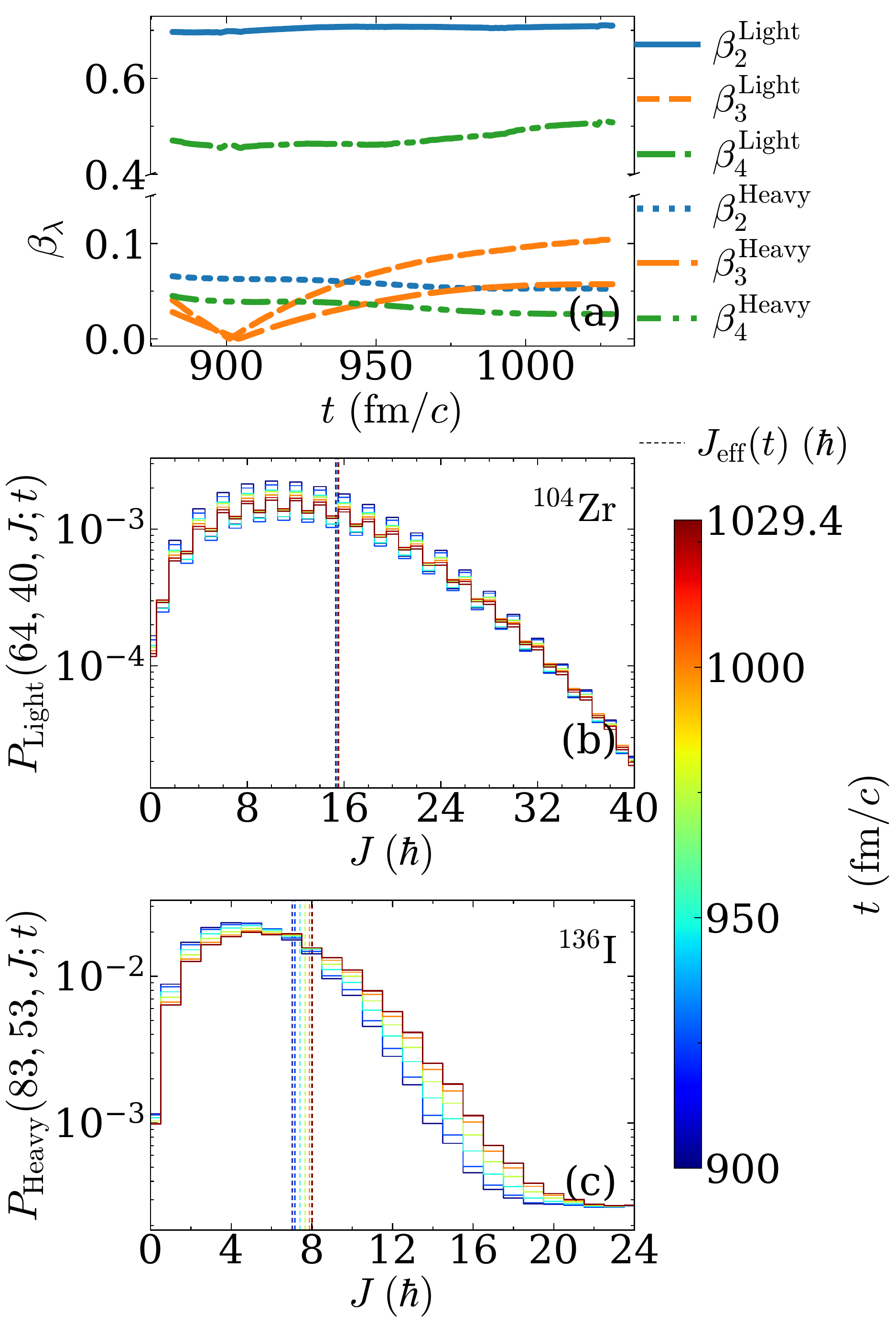}
  \caption{Time evolution of the fission-fragment deformation parameters and angular-momentum distributions. Panel (a) shows the quadrupole, octupole, and hexadecapole deformation parameters of the light and heavy fragments. Panels (b) and (c) show the angular-momentum distributions and its expectation values for the light-fragment channel $(N,Z)=(64,40)$ and the heavy-fragment channel $(N,Z)=(83,53)$, respectively.}
  \label{fig:beta234_amp_fission}
\end{figure}

Finally, we examine the time dependence of the angular-momentum distributions in
induced fission, in a similar mannar as in Sec.~\ref{sec:td-dist-MNT}.
Figure~\ref{fig:beta234_amp_fission}(a) shows the time evolution of the deformation
parameters, $\beta_2$, $\beta_3$, and $\beta_4$, of the lighter and heavier fragments.
As noted above, the lighter fragment remains strongly deformed, with $\beta_2^{\mathrm{light}}
\sim0.7$, and its deformation changes little with time. In Figs.~\ref{fig:beta234_amp_fission}(b)
and \ref{fig:beta234_amp_fission}(c), we show the angular-momentum distributions of
a lighter and a heavier fragments, $^{104}$Zr and $^{136}$I, respectively. Again,
the PNP+AMP calculation is performed at every $5$\,fm/$c$, and corresponding
times are indicated by line colors. In Figs.~\ref{fig:beta234_amp_fission}(b)
and \ref{fig:beta234_amp_fission}(c), we find a spin-cutoff-like distribution for
both of the lighter and heavier fragments. In the case of $^{104}$Zr (a well
deformed lighter fragment) shown in Fig.~\ref{fig:beta234_amp_fission}(b),
we also find a odd-even staggering associated with a rotational band of the
largely deformed lighter fragment, while its mean angular momentum (indicated
by vertical dotted lines) remains nearly constant. As time evolves, the odd-even
staggering becomes smaller, which may be suggesting possible couplings between
collevtive and single-particle excitations though the one-body dissipation mechanism.
On the other hand, in the case of $^{136}$I (weakly deformed heavier fragment),
we observe that a sift of the mean angular momentum (indicated by vertical
dotted lines) from $J\simeq 7\hbar$ to $8\hbar$, indicating an increase of
high-spin components in time after scission.

Except for such a time dependence, the angular-momentum distributions of the
light and heavy fragments are similar to those observed in the multinucleon
transfer reaction [cf.\ Figs.~\ref{fig:beta234_amp_fission}(b) and \ref{fig:beta234_amp_fission}(c)].
Although the distributions vary with time, our results indicate that the
corresponding probabilities remain within the same order of magnitude.
These variations are therefore unlikely to alter the qualitative interpretation,
particularly with regard to the overall shapes of the distributions.
For quantitative comparisons, however, the increase in the mean angular momentum
observed in Fig.~\ref{fig:beta234_amp_fission}(c) should be taken into consideration.

\section{Summary and prospect}\label{Sec:Conclusion}

In this work, we have investigated the angular-momentum distributions of
primary reaction products produced though multinucleon transfer processes
in the $^{40}\mathrm{Ca}+^{124}\mathrm{Sn}$ reaction based on the time-dependent
Hartree-Fock (TDHF) theory combined with the particle-number and angular-momentum
projections. The simultaneous particle-number and angular-momentum projections
(PNP+AMP) analysis enables us to calculate the angular-momentum distributions,
$P(N,Z,J)$, for each transfer channels from the TDHF wave functions after collision.
This framework provides microscopic information on fragment angular momentum
that is inaccessible from either PNP or AMP alone.

For one-nucleon-transfer channels, we have found characteristic peaks in
the calculated angular-momentum distributions which can be associated with
the single-particle orbitals involved in the transfer process. The distributions
therefore retain clear signatures of the shell structure in the reaction product.
For two-nucleon-transfer channels, we have reconstructed the angular-momentum
distributions by coupling the corresponding one-nucleon-transfer distributions
using the Clebsch--Gordan coupling probabilities supplemented with the Pauli
selection rule. The reconstructed distributions reproduce fairly well the main
features of the fully microscopic TDHF+PNP+AMP results, indicating that the
angular momenta in these channels can still be understood, to a good approximation,
in terms of combinations of single-particle angular momenta.

As the number of transferred nucleons increases, distributions become smoother
and gradually resemble the so-called spin-cutoff form which assumes statistical
equilibration. The growing agreement with this spin-cutoff form with an increasing
number of trasnfered nucleons suggests that the expanding number of angular-momentum-coupling
configurations and the increasing complexity of multiparticle--multihole excitations
obscure the individual single-particle structures. By fitting the microscopic
distributions with the spin-cutoff form, we have extracted the spin-cutoff
parameter and compared the resulting effective angular momenta and distribution
widths with those obtained from the empirical prescription employed in a statistical
model, CGMF. For channels involving larger numbers of transferred nucleons,
the TDHF distributions yield larger average angular momenta and broader widths,
indicating that the empirical prescription may underestimate the angular momentum
carried by the primary fragments in the reaction considered here.

For completeness to validate the proposed PNP+AMP method, we have also examined
the time dependence of the calculated distributions. Before the PLF and TLF are
sufficiently separated, the distributions depend appreciably on the definition
of the fragment region. After reseparation, on the other hand, their principal
structures remain nearly unchanged despite the continuing evolution of the
fragment shapes. The interpretations based on the single-particle peaks and
spin-cutoff-like distributions are therefore insensitive to the precise time
at which the PNP and AMP projections are performed, provided that the fragments
are sufficiently separated.

As an exploratory application, we applied the same PNP+AMP analysis to fission
fragments generated through induced fission of $^{240}\mathrm{Pu}$. The calculated
angular-momentum distributions of fission fragments exhibit spin-cutoff-like components
as well as rotational-band-like pattern associated with a large deformation of
lighter fragments, influenced with single-particle shell structure. In particular,
the angular-momentum distributions of fragments formed near the magic numbers
show pronounced deviations from the spin-cutoff form. These results demonstrate
that the present method can be used to distinguish different microscopic contributions
to fragment angular momentum in both multinucleon transfer and fission processes.

Amongst the huge body of possible extensions of the present approach, we briefly
mention here some of them to stimulate future works. Concerning the projection
analysis, we have carried out simultaneous PNP and AMP to determin angular-momentum
distributions in each fragment. One can, in principle, perform the parity
projection as well, which will allow us explicit spin-pairty assignments
for excited states populated via the multinucleon transfer reaction. By calculating
the energy expectation value of such a fully projected state, we can not only
construct a level scheme of the reaction product, but also calculate transitions
between these excited states. Through this kind of analysis, we can make a
direct comparison between microscopic calculations with experimental data
of $\gamma$-coinsidence measurements for multinucleon transfer products.
In addition, the angular-momentum distributions and excitation energies
are important quantity for evaluating secondary disintegration processes
through particle evapolation or fission. Since significantly wider spin
distributions apart from the mean value were observed in the PNP+AMP
analysis, the use of those microscopically detemined $P(N,Z,J^\pi,E^*)$
as imputs of statistical model calculations for secondary decays may
significantly alter final yields. Last but not least, it goes without
saying that the pairing correlations would play an important role in
dynamics in reactions and fission, as was partly discussed in the main
texts. Although it requires enomous computational effort to carry out
fully microscopic dynamic simulations including pairing correlations
combined with the PNP+AMP(+parity projection) analysis, it could be
a challenge at the frontier of computational nuclear science with
top-tier Exascale supercomputers.

In conclusion, this work has elucidated the richness of physics contained already
in the standard TDHF theory and the feasibility of calculating angular-momentum
distributions through the combined PNP+AMP analysis. This way, this work opens
a door to new research posibilities where microscopically-calculated spin-resolved
reaction observables can be compared directly with experimental data.
Detailed quantitative comparisons of such spin-resolved new observables
may be useful to constrain yet-well-determined components in the energy
density functional such as spin-dependent time-odd terms and/or tensor
terms.

\section*{Acknowledgments}
We would like to express our gratitude to Mr.\ Shu Yamamura (Institute
of Science Tokyo) for valuable discussions. We would also like to express
our gratitude to Dr.\ Kotaro Uzawa (JAEA) for stimulating discussion that
triggered this work. This work is supported by JSPS Grant-in-Aid for
Scientific Research, Grants No.~JP25H01269. This research was supported
by the Science Tokyo Support Program for Doctoral Students, funded by
the Universities for International Research Excellence. This work used
computational resources of the Heian supercomputer at Yukawa Institute
for Theoretical Physics (YITP), Kyoto University. This work also used
(in part) the computational resources of Miyabi at University of Tokyo,
through the HPCI System Project, Project ID: hp260222.


\bibliography{cite}



\section*{Supplementary material}

# Supplemental Material: Microscopic description of spin distributions in multinucleon transfer reactions

Kotaro Koga[1,*] and Kazuyuki Sekizawa[1,2,3,†]

[1]Department of Physics, School of Science, Institute of Science Tokyo, Tokyo 152-8550, Japan
[2]Nuclear Physics Division, Center for Computational Sciences, University of Tsukuba, Ibaraki 305-8577, Japan
[3]RIKEN Nishina Center, Saitama 351-0198, Japan



## I. SINGLE-PARTICLE LEVELS

Figures 1 and 2 show the single-particle levels of $^{40}$Ca and $^{132}$Sn used in the discussion below. The vertical axis gives the single-particle energy, while the left and right columns show neutron and proton levels, respectively. Solid lines denote occupied orbitals, and dashed lines denote unoccupied orbitals. Positive-energy continuum states are omitted because the angular-momentum expectation values of these discretized states do not represent well-defined single-particle quantum numbers.

## II. ONE-NUCLEON-TRANSFER CHANNELS

Figure 3 shows the angular-momentum distributions for the $(0n,-1p)$ and $(0n,+1p)$ channels, which are not shown in the main text. The plotting conventions are the same as in Fig. 3 of the main text. The $(0n,-1p)$ channel exhibits a distribution similar to that of the $(-1n,0p)$ channel, whereas the $(0n,+1p)$ channel has a longer tail extending beyond $J=11/2\hbar$. As indicated by the single-particle levels in Fig. 1, this difference arises because the relevant proton states lie lower in energy than the corresponding neutron states, while the levels above the $1f_{7/2}$ orbital belong to the continuum.

## III. TWO-NUCLEON-TRANSFER CHANNELS

Figures 4, 5, 6, and 7 show the angular-momentum distributions for the $(\pm 1n,\pm 1p)$ channels omitted from the main text. In addition to the total angular-momentum distribution [panel (c) of each figure], panels (a) and (b) show the neutron and proton angular-momentum distributions, $P(J_N \mid N,Z)$ and $P(J_Z \mid N,Z)$, respectively. These distributions are obtained by applying Eq. (18) of the main text separately to neutrons and protons:

$$P(N_q,J_q) = \frac{2J_q+1}{16\pi^2}\int \chi^*_{J_q}(\Omega)K^{(q)}(N_q,\Omega)\,\mathrm{d}\Omega, \qquad (1)$$

Equation 1 is evaluated for $q=\mathrm{n},\mathrm{p}$. The neutron and proton distributions conditioned on the channel $(N,Z)$ are constructed by multiplying $P(N,J_N)$ and $P(Z,J_Z)$ by $P(Z)$ and $P(N)$, respectively, and then normalizing the resulting distributions. The total angular momentum $J$ is obtained by coupling $J_N$ and $J_Z$ using the product rule for Wigner $D$ functions. Panels (a) and (b) may be interpreted as the neutron and proton one-nucleon-transfer components, respectively, and therefore exhibit shell effects similar to those found in the one-nucleon-transfer channels. The total distribution in panel (c) is obtained by coupling these two angular momenta. This coupling is performed exactly through angular-momentum projection rather than with the approximate coupling model introduced in the main text.

## IV. EXCITATION ENERGIES USED IN THE CGMF CALCULATIONS

Table I summarizes the quantities used to evaluate the CGMF spin-cutoff parameter $B_{\mathrm{CGMF}}$ and the values of $B_{\mathrm{TDHF}}$ obtained by fitting the microscopic angular-momentum distributions. The quantities listed in the table are defined in the main text.

* koga.k.c19d@m.isct.ac.jp

† sekizawa@phys.sci.isct.ac.jp

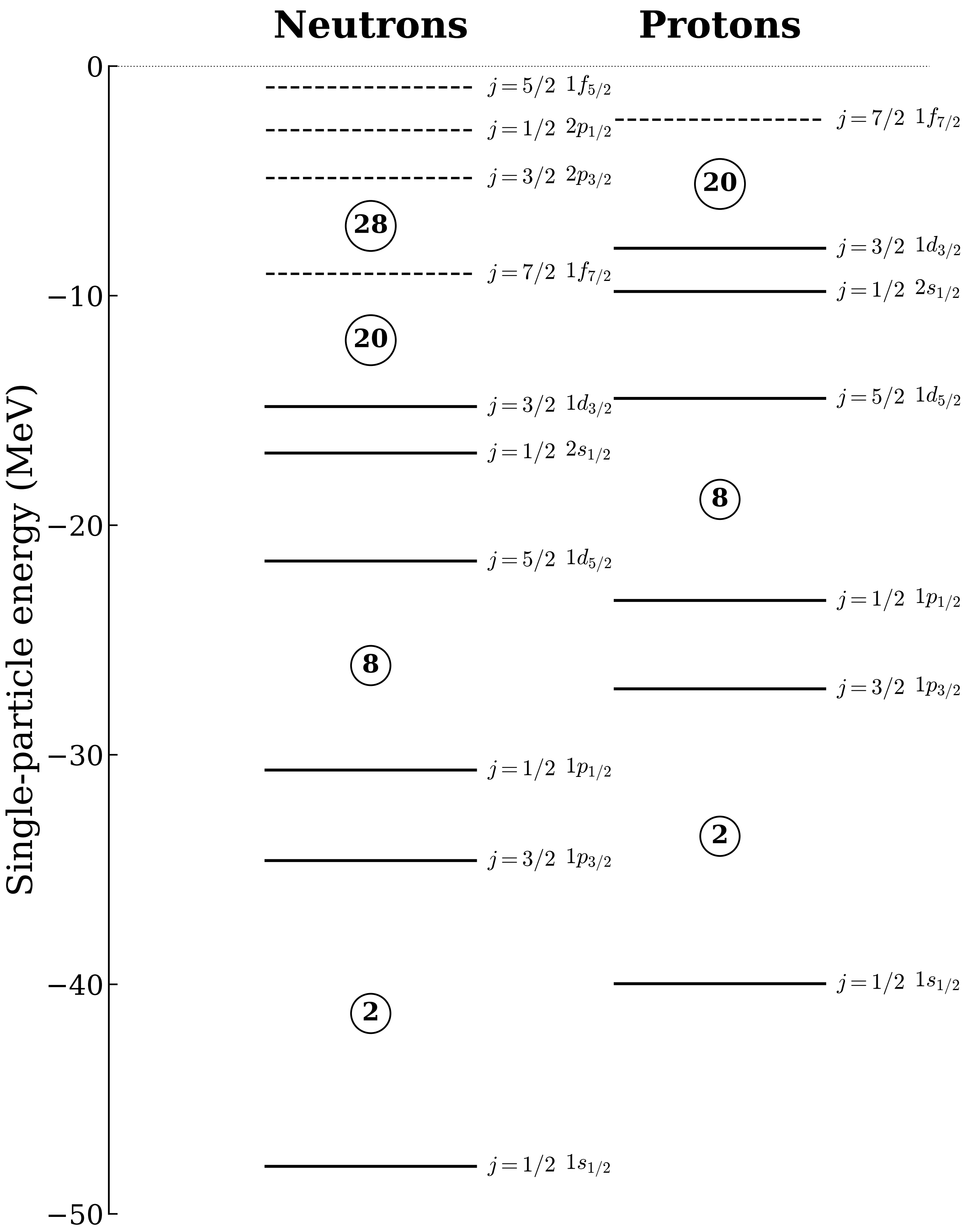


FIG. 1. Single-particle energy levels of $^{40}$Ca. Neutron and proton levels are shown in the left and right columns, respectively. Solid (dashed) lines denote occupied (unoccupied) orbitals; positive-energy continuum states are omitted.

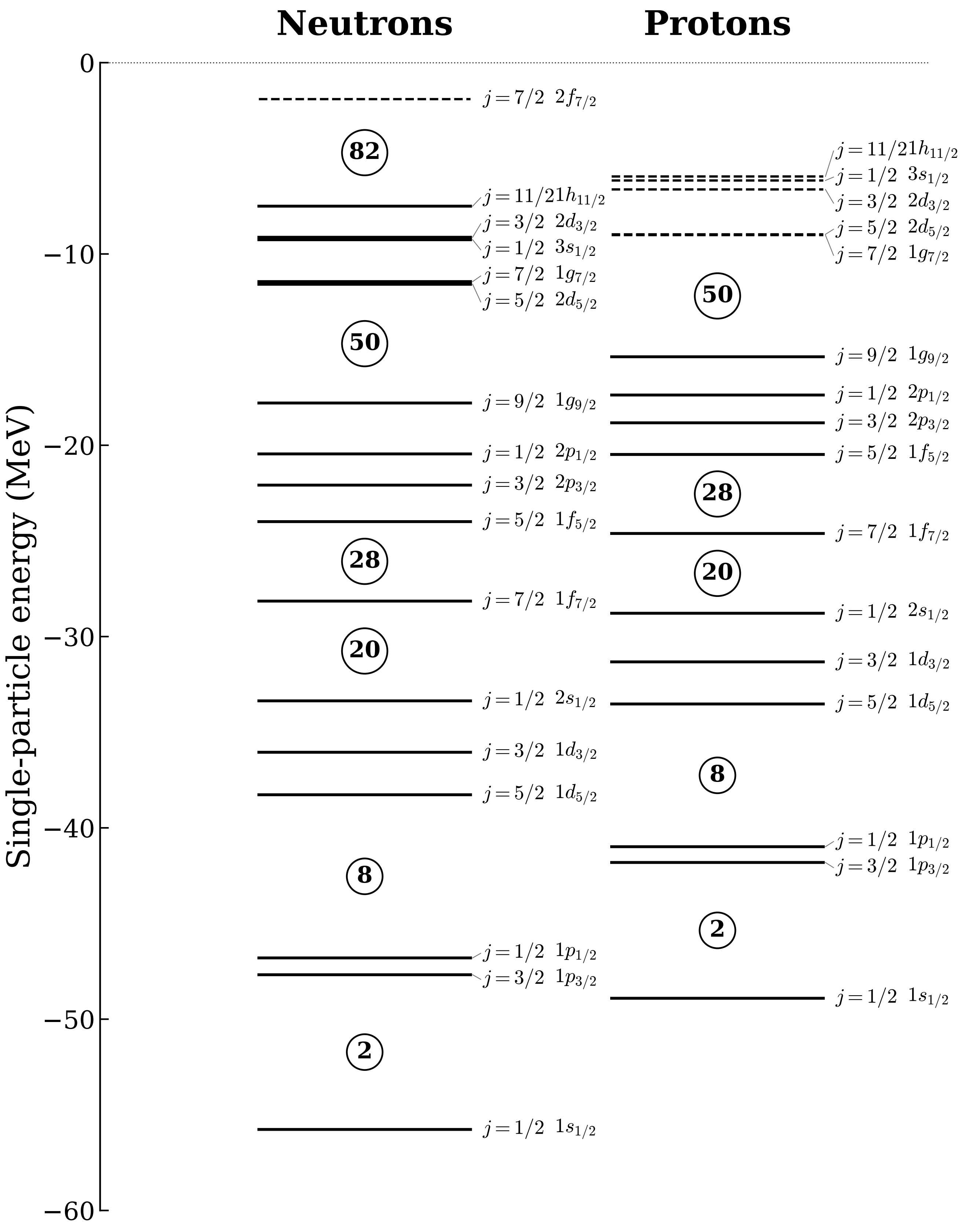


FIG. 2. Single-particle energy levels of $^{132}$Sn. Neutron and proton levels are shown in the left and right columns, respectively. Solid (dashed) lines denote occupied (unoccupied) orbitals; positive-energy continuum states are omitted.

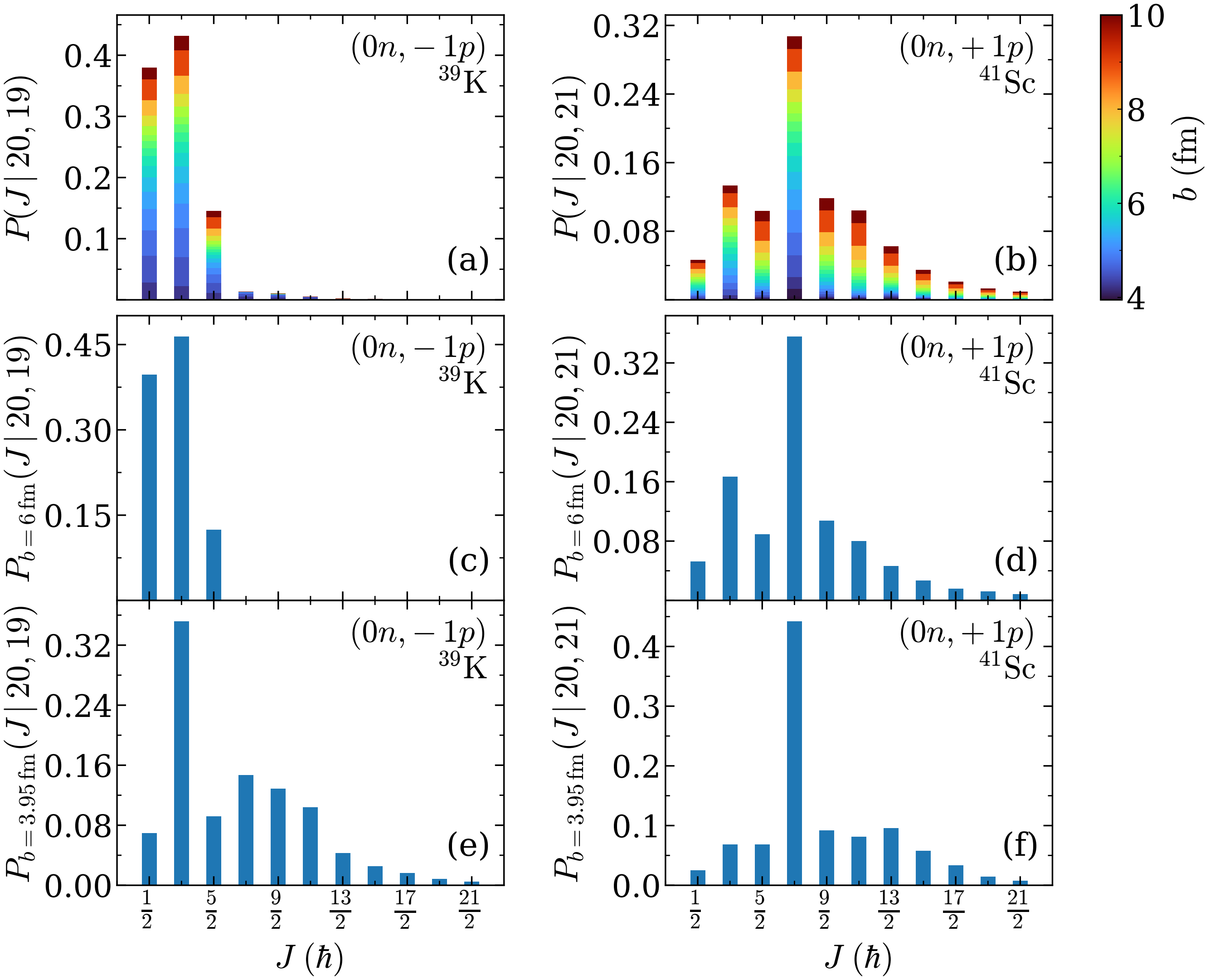


FIG. 3. Angular-momentum distributions for the one-proton-removal and one-proton-addition channels of the PLF. Panels (a), (c), and (e) show the $(0n,-1p)$ channel, whereas panels (b), (d), and (f) show the $(0n,+1p)$ channel. Panels (a) and (b) give the impact-parameter-integrated distributions; panels (c) and (d) and panels (e) and (f) give the distributions at $b = 6.00$ and $3.95\,\mathrm{fm}$, respectively.

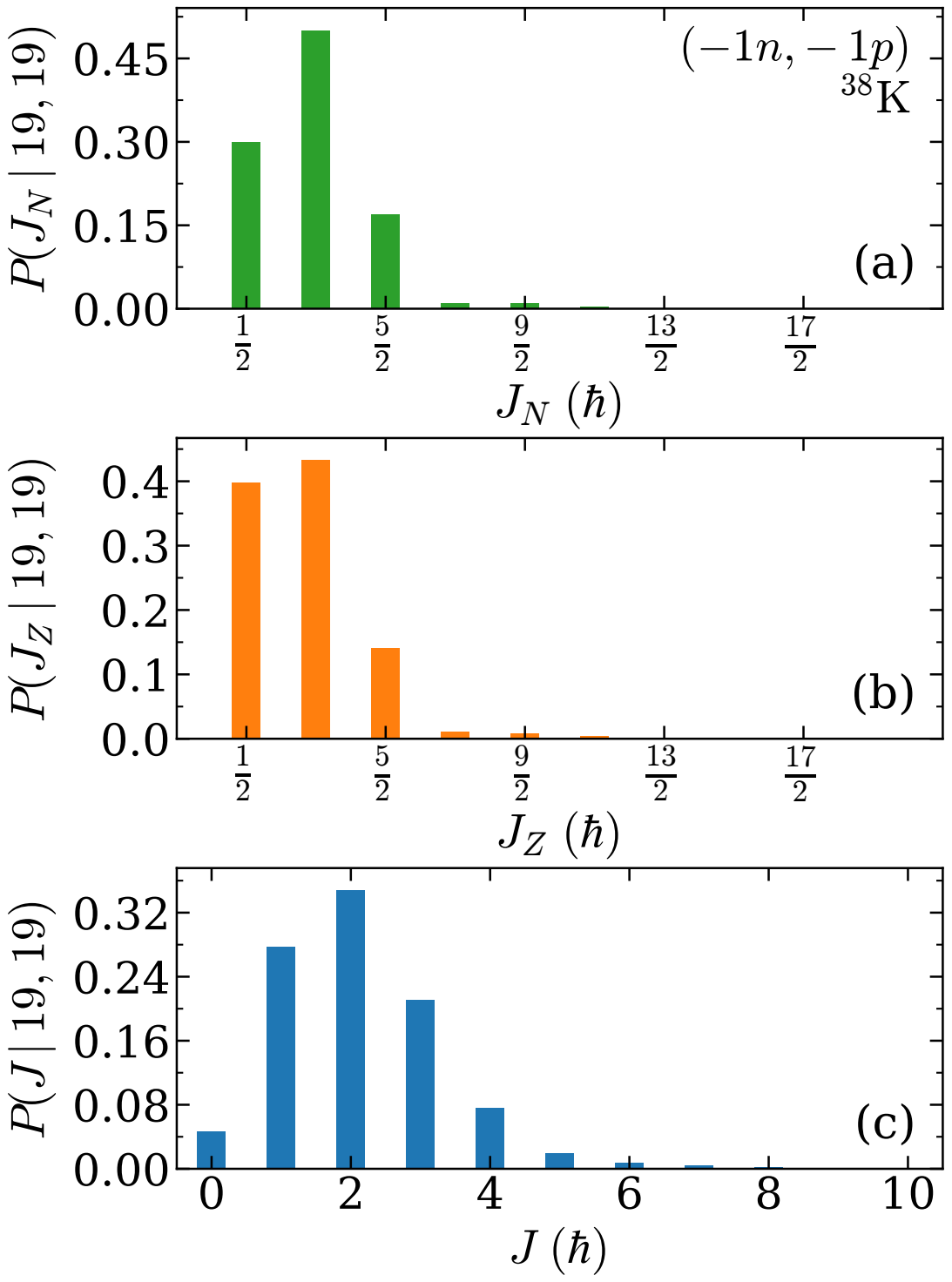


FIG. 4. Neutron, proton, and total angular-momentum distributions of the PLF for the $(-1n, -1p)$ channel. Panels (a), (b), and (c) show $P(J_N \mid N, Z)$, $P(J_Z \mid N, Z)$, and $P(J \mid N, Z)$, respectively.

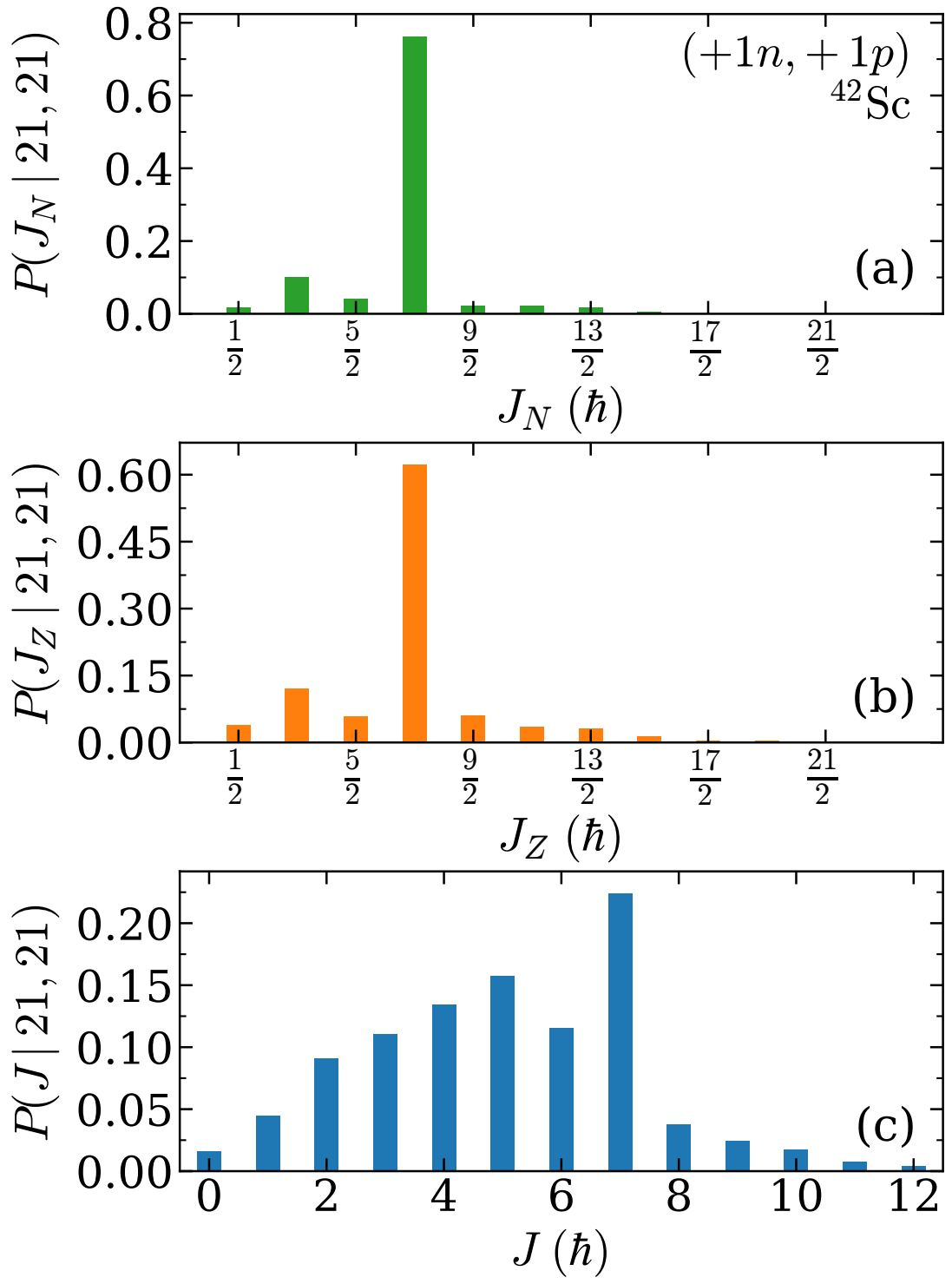


FIG. 5. Neutron, proton, and total angular-momentum distributions of the PLF for the $(+1n, +1p)$ channel. Panels (a), (b), and (c) show $P(J_N \mid N, Z)$, $P(J_Z \mid N, Z)$, and $P(J \mid N, Z)$, respectively.

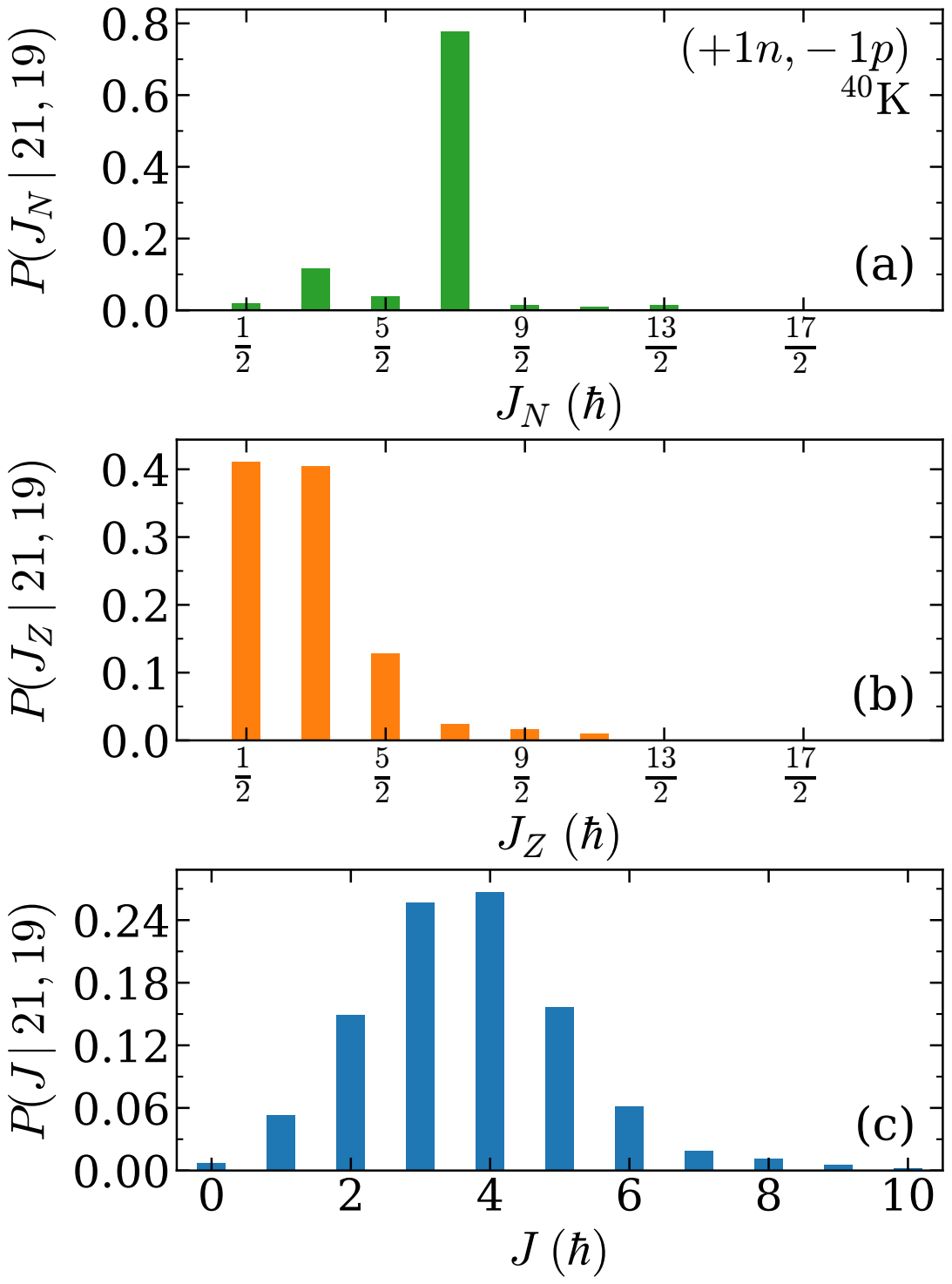


FIG. 6. Neutron, proton, and total angular-momentum distributions of the PLF for the $(+1n, -1p)$ channel. Panels (a), (b), and (c) show $P(J_N \mid N, Z)$, $P(J_Z \mid N, Z)$, and $P(J \mid N, Z)$, respectively.

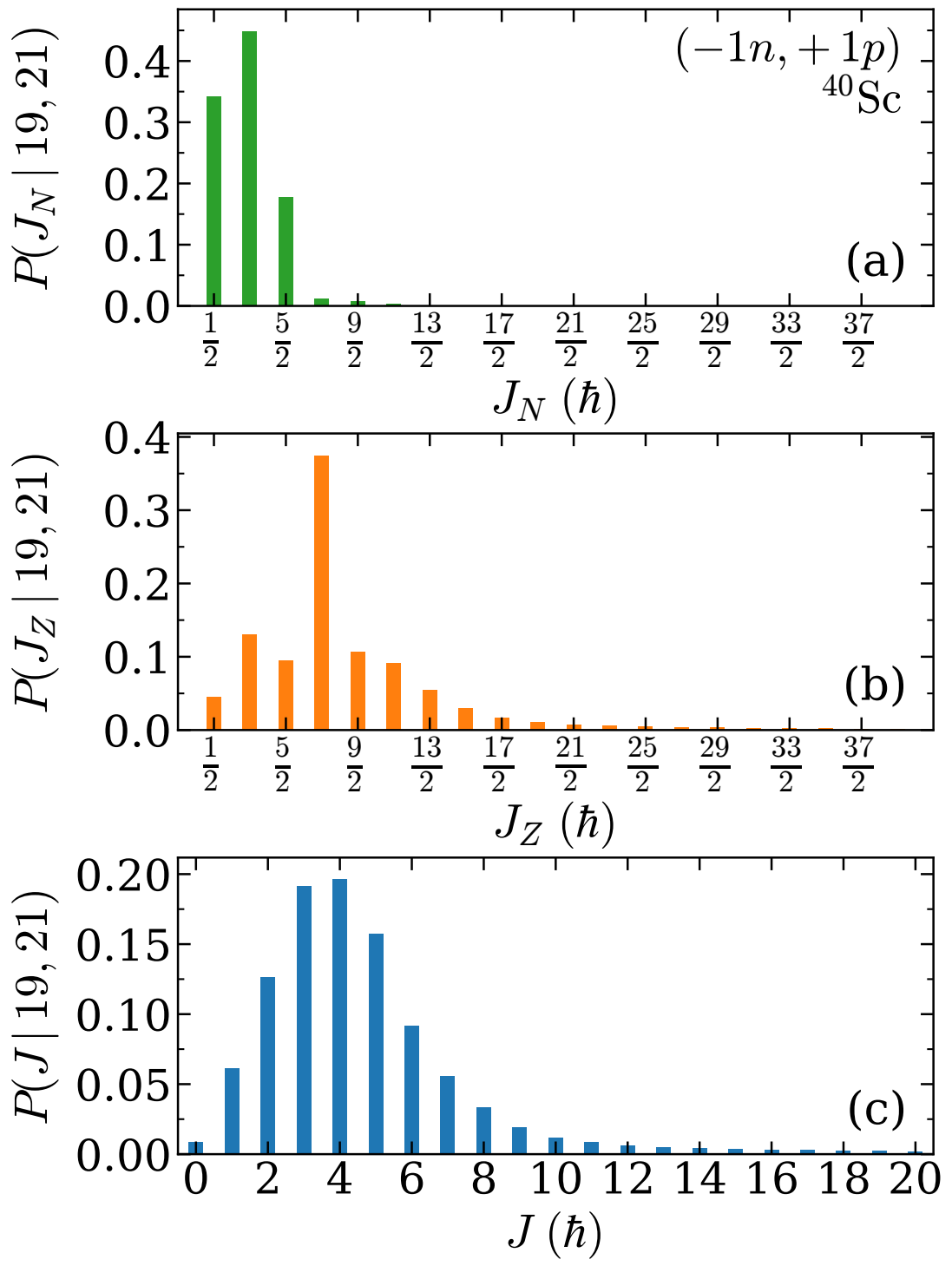


FIG. 7. Neutron, proton, and total angular-momentum distributions of the PLF for the $(-1n, +1p)$ channel. Panels (a), (b), and (c) show $P(J_N \mid N, Z)$, $P(J_Z \mid N, Z)$, and $P(J \mid N, Z)$, respectively.

TABLE I. Quantities used to determine $B_{\rm CGMF}$ and results of fitting the microscopic angular-momentum distributions. Listed are the transferred-nucleon number $A_{\rm tr}$, neutron and proton numbers $(N,Z)$, channel probability $P(N,Z)$, excitation energy $E^*$, CGMF spin-cutoff parameter $B_{\rm CGMF}$, fitted value $B_{\rm TDHF}$, and coefficient of determination $R^2$.

| $A_{\rm tr}$ | $N$ | $Z$ | $P(N,Z)$ | $E^*$ (MeV) | $B_{\rm CGMF}$ | $B_{\rm TDHF}$ | $R^2$ |
|---|---|---|---|---|---|---|---|
| 0 | 20 | 20 | $4.941401\times10^{-3}$ | 6.494710 | 3.186002 | 3.449379 | 0.415765 |
| 1 | 19 | 20 | $3.327592\times10^{-4}$ | 3.977715 | 2.979233 | 2.854424 | 0.870516 |
| 1 | 20 | 19 | $1.733249\times10^{-2}$ | 6.090831 | 3.160780 | 2.660266 | 0.721077 |
| 1 | 20 | 21 | $2.922510\times10^{-4}$ | 3.905457 | 3.307770 | 3.760587 | 0.526413 |
| 1 | 21 | 20 | $1.833790\times10^{-2}$ | 6.625757 | 3.441272 | 3.857163 | 0.533295 |
| 2 | 19 | 19 | $1.167437\times10^{-3}$ | 4.343199 | 2.891712 | 2.798972 | 0.959049 |
| 2 | 20 | 18 | $2.389596\times10^{-2}$ | 6.564136 | 3.247003 | 2.763897 | 0.719895 |
| 2 | 21 | 19 | $6.433714\times10^{-2}$ | 6.025610 | 3.203319 | 3.932267 | 0.851636 |
| 2 | 21 | 21 | $1.084821\times10^{-3}$ | 4.928034 | 3.347586 | 4.805280 | 0.811774 |
| 2 | 22 | 20 | $2.771204\times10^{-2}$ | 8.204772 | 3.573762 | 4.322312 | 0.800625 |
| 3 | 19 | 18 | $1.609206\times10^{-3}$ | 5.139077 | 3.179321 | 2.833502 | 0.970680 |
| 3 | 20 | 17 | $1.461520\times10^{-2}$ | 5.481235 | 3.161746 | 2.597002 | 0.891199 |
| 3 | 21 | 18 | $8.868200\times10^{-2}$ | 6.138796 | 3.361407 | 3.957401 | 0.843298 |
| 3 | 22 | 19 | $9.720211\times10^{-2}$ | 7.083219 | 3.556140 | 4.306247 | 0.962153 |
| 3 | 22 | 21 | $1.638934\times10^{-3}$ | 6.705163 | 3.664479 | 4.933560 | 0.926537 |
| 3 | 23 | 20 | $1.848664\times10^{-2}$ | 8.168934 | 3.657391 | 4.116659 | 0.917854 |
| 4 | 19 | 17 | $9.844052\times10^{-4}$ | 4.568153 | 2.968401 | 2.786541 | 0.962905 |
| 4 | 20 | 16 | $3.974664\times10^{-3}$ | 5.284780 | 3.088318 | 2.683259 | 0.792942 |
| 4 | 21 | 17 | $5.425068\times10^{-2}$ | 4.925877 | 2.934769 | 3.942014 | 0.870999 |
| 4 | 22 | 18 | $1.340140\times10^{-1}$ | 7.101092 | 3.498622 | 4.307494 | 0.925758 |
| 4 | 23 | 19 | $6.486035\times10^{-2}$ | 6.890733 | 3.559830 | 4.162041 | 0.957572 |
| 4 | 23 | 21 | $1.093629\times10^{-3}$ | 7.269179 | 3.650164 | 4.863407 | 0.955938 |
| 4 | 24 | 20 | $4.992058\times10^{-3}$ | 9.689902 | 3.646782 | 4.046946 | 0.865107 |
| 5 | 19 | 16 | $2.676646\times10^{-4}$ | 4.501494 | 3.122917 | 2.742962 | 0.979114 |
| 5 | 21 | 16 | $1.475087\times10^{-2}$ | 4.234186 | 3.254162 | 3.903804 | 0.827003 |
| 5 | 22 | 17 | $8.196425\times10^{-2}$ | 5.276100 | 3.366071 | 4.269094 | 0.964855 |
| 5 | 23 | 18 | $8.940152\times10^{-2}$ | 6.447231 | 3.411561 | 4.162231 | 0.952683 |
| 5 | 24 | 19 | $1.750958\times10^{-2}$ | 7.793747 | 3.592172 | 4.005054 | 0.976555 |
| 5 | 24 | 21 | $2.952284\times10^{-4}$ | 8.868681 | 3.811090 | 4.750024 | 0.899782 |
| 5 | 25 | 20 | $4.543326\times10^{-4}$ | 9.446612 | 3.748370 | 4.235689 | 0.955075 |
| 6 | 22 | 16 | $2.229108\times10^{-2}$ | 4.531741 | 3.328536 | 4.234129 | 0.903864 |
| 6 | 23 | 17 | $5.469220\times10^{-2}$ | 4.505563 | 3.492803 | 4.127251 | 0.960895 |
| 6 | 24 | 18 | $2.414125\times10^{-2}$ | 7.244077 | 3.369316 | 4.009391 | 0.942571 |
| 6 | 25 | 19 | $1.594050\times10^{-3}$ | 7.444872 | 3.563756 | 4.257690 | 0.960381 |
| 7 | 23 | 16 | $1.487062\times10^{-2}$ | 3.308453 | 3.501493 | 4.076307 | 0.952044 |
| 7 | 24 | 17 | $1.476472\times10^{-2}$ | 4.700054 | 3.571008 | 3.971562 | 0.979483 |
| 7 | 25 | 18 | $2.197156\times10^{-3}$ | 6.319783 | 3.547582 | 4.253233 | 0.956277 |
| 8 | 24 | 16 | $4.015524\times10^{-3}$ | 3.428217 | 3.598829 | 3.944955 | 0.913058 |
| 8 | 25 | 17 | $1.344155\times10^{-3}$ | 3.703552 | 3.658311 | 4.218836 | 0.956130 |
| 9 | 25 | 16 | $3.654661\times10^{-4}$ | 1.951818 | 3.664960 | 4.172256 | 0.953919 |



\end{document}